\documentclass[prb, aps, showpacs,superscriptaddress, twocolumn, longbibliography]{revtex4-2}
\usepackage[colorlinks=true,linkcolor=blue,urlcolor=blue,citecolor=blue]{hyperref}

\usepackage[utf8]{inputenc}
\usepackage{color}
\usepackage{bbm} 

\usepackage{amsfonts,amsmath,amssymb,stmaryrd}

\usepackage{graphicx}
\usepackage{subfigure}  
\usepackage{bbm} 
\usepackage{hyperref}
\usepackage{epsfig}
\usepackage{mathrsfs}
\usepackage{verbatim}
\usepackage{centernot}
\usepackage{ulem}
\usepackage{longtable}
\usepackage{nicefrac}
\usepackage[framemethod=tikz]{mdframed}

\usepackage{pdfpages}
\usepackage{etoolbox} 
\makeatletter
\patchcmd{\@outputpage@head}{\@ifx{\LS@rot\@undefined}{}{\LS@rot}}{}{}{}
\makeatother

\begin{document}

\title{Magnetism in the two-dimensional dipolar XY model}

\author{Björn Sbierski}
\affiliation{Department of Physics and Arnold Sommerfeld Center for Theoretical Physics (ASC), Ludwig-Maximilians-Universit\"at M\"unchen, Theresienstr.~37, M\"unchen D-80333, Germany}
\affiliation{Munich Center for Quantum Science and Technology (MCQST), Schellingstr.~4, D-80799 M\"unchen, Germany}

\author{Marcus Bintz}
\affiliation{Department of Physics, Harvard University, Cambridge, Massachusetts 02138 USA}

\author{Shubhayu Chatterjee}
\affiliation{Department of Physics, Carnegie Mellon University, Pittsburgh, Pennsylvania 15213, USA}
\affiliation{Department of Physics, University of California, Berkeley, California 94720, USA}

\author{Michael Schuler}
\affiliation{Institut für Theoretische Physik, Universität Innsbruck, A-6020 Innsbruck, Austria}

\author{Norman Y. Yao}
\affiliation{Department of Physics, Harvard University, Cambridge, Massachusetts 02138 USA}
\affiliation{Department of Physics, University of California, Berkeley, California 94720, USA}
\affiliation{Material Science Division, Lawrence Berkeley National Laboratory, Berkeley, California 94720, USA}

\author{Lode Pollet}
\affiliation{Department of Physics and Arnold Sommerfeld Center for Theoretical Physics (ASC), Ludwig-Maximilians-Universit\"at M\"unchen, Theresienstr.~37, M\"unchen D-80333, Germany}
\affiliation{Munich Center for Quantum Science and Technology (MCQST), Schellingstr.~4, D-80799 M\"unchen, Germany}

\date{\today}

\begin{abstract}
Motivated by a recent experiment on a square-lattice Rydberg atom array realizing a long-range dipolar XY model [Chen {\it et al.}, Nature (2023)], we numerically study the model's equilibrium properties. We obtain the phase diagram, critical properties, entropies, variance of the magnetization, and site-resolved correlation functions. We consider both ferromagnetic and antiferromagnetic interactions and apply quantum Monte Carlo and pseudo-Majorana functional renormalization group techniques, generalizing the latter to a $U(1)$ symmetric setting.
Our simulations perform extensive thermometry for the first time in dipolar Rydberg atom arrays and establish conditions for adiabaticity and thermodynamic equilibrium. On the ferromagnetic side of the experiment, we determine the entropy per particle $S/N \approx 0.5$, close to the one at the critical temperature, $S_c/N = 0.585(15)$. The simulations suggest the presence of an out-of-equilibrium plateau at large distances in the correlation function, thus motivating future studies on the non-equilibrium dynamics of the system.
\end{abstract}

\maketitle

\section{Introduction}

Quantum many-body systems with long-range interactions can harbor richer physics than systems with short-range interactions. Prominent examples are Wigner crystals~\cite{Li_Wigner2021, Smolenski_Wigner2021}, exotic superconductivity in magic-angle twisted bilayer graphene~\cite{Cao2018, Cao2018b}, the fractional quantum Hall effect~\cite{Tsui_FQHE1982}, trapped ions~\cite{Blatt2012, Richerme2014, Jurcevic2014, MonroeRMP2021, feng2022continuous}, and atoms coupled to cavities~\cite{Ritsch_RMP2013}. 
Dipolar interactions in particular may stabilize several strongly correlated phases such as supersolids~\cite{Pollet_supersolid_2010, Capogrosso_supersolid_2010}, the Haldane phase~\cite{DallaTorre2006}, and spin ice~\cite{castelnovoMagneticMonopolesSpin2008}.
In these systems competing interactions remain poorly understood and quantum simulation of long-range interactions, and dipolar interactions in particular, is therefore highly called for.  Although progress in recent years has been rapid for polar molecules in optical lattices~\cite{Moses2017, Bohn2017, DeMarco2019, Schindewolf2022, Capogrosso2015, Capogrosso2018, Capogrosso2022, Christakis2023}, we focus here on the realization of dipolar interactions with two-dimensional Rydberg atom arrays~\cite{Morgado2021, Henriet2020quantumcomputing} which currently allow for higher filling fractions.
\begin{figure}
    \centering
    \includegraphics[width=\columnwidth]{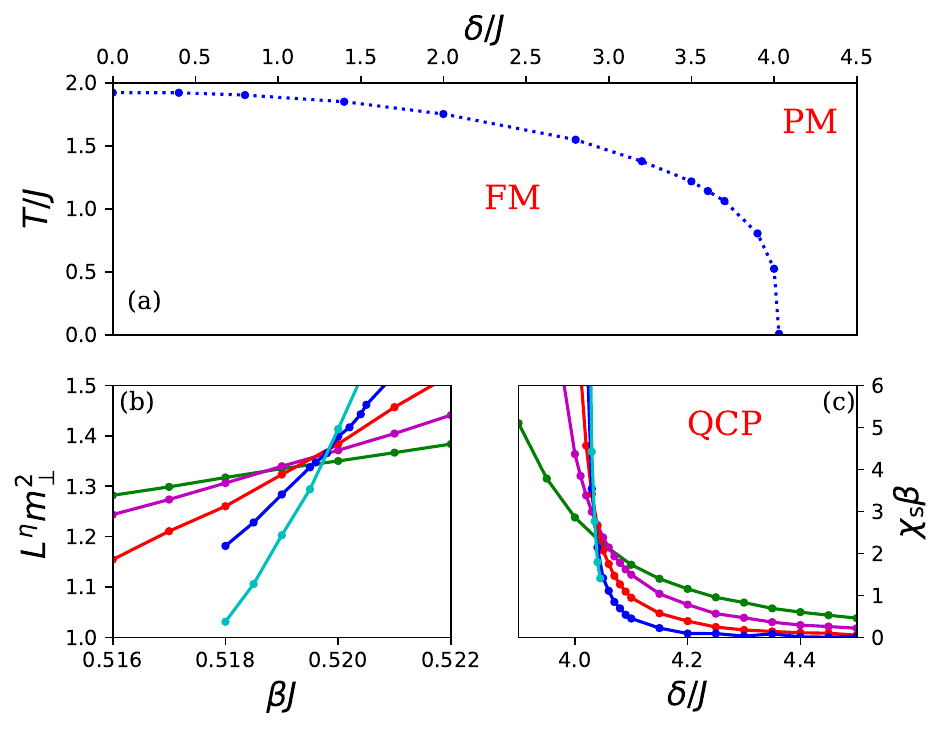}
    \caption{FM case: (a) Phase diagram for the $XY$ model with FM dipolar interactions and a staggered z-field $\delta$, see Eq.~\eqref{eq:Hxy+Hz}. 
    The dotted line is a guide to the eye separating the paramagnetic phase (PM) and in-plane $XY$ FM order. (b) Finite-T transition: In-plane magnetization squared $m_{\perp}^2$ multiplied by $L^{\eta}$ with $\eta=1$ as a function of inverse temperature $\beta$ for linear system sizes $L = 17, 33, 65, 127, 257$ (with increasing slope) for the model without staggered field, $\delta = 0$. (c) Quantum critical point (QCP): 
    Finite-size scaling  of the spin stiffness $\chi_{\rm s}$ multiplied by the inverse temperature $\beta$ as a function of staggered field $\delta$ for system sizes as in (b) and scaling the inverse temperature as $\beta \sim \sqrt{L}$ initiated by $\beta(L=33)=16$. Error bars are smaller than marker size in all subplots.}
     \label{fig:QMC_scaling}
\end{figure}
Experimentally, the dipolar XY-model for spin $S=1/2$ emerges as the effective Hamiltonian for atomic tweezer arrays where the effective spin is encoded in a pair of highly excited Rydberg states; the model's $U(1)$ symmetry is inherited from energy conservation, which then leads to conservation of excitation number~\cite{Barredo2014,Leseleuc2019,Lienhard2020}.
Despite the continuous $U(1)$ symmetry, the long-range nature of this interaction (power-law decay with exponent $\alpha=3$) evades the Mermin-Wagner theorem in two dimensions which applies only for $\alpha \ge 4$ \cite{Mermin1966, Bruno2001}. This allows for true long-range order at $T>0$. In terms of spin waves, power-law ferromagnetic (FM) interactions cause the low-energy dispersion to behave as $\omega_k \sim k^{(\alpha-2)/2}$. For $2<\alpha<4$, the $k=0$ singularity suppresses the density of spin waves and their proliferation at small $T$ \cite{Peter2012, diesselGeneralizedHiggs2022,Song2023}. Such systems have interesting properties  for quantum metrology and sensing~\cite{NormYao_squeezing_2023, Comparin2022, roscildeEntanglingDynamics2023}. For antiferromagnetic (AFM) dipolar interactions, frustration effects maintain a linear spin-wave dispersion as in the short-range case, and thus prevent long-range order at $T>0$.

Progress in experimental techniques for large-scale Rydberg atom arrays \cite{Schauss2015,Ebadi2021,Semeghini2021,Browaeys2020} has allowed testing the above predictions in the laboratory: A recent experiment realized the dipolar XY model on a two-dimensional square lattice consisting of up to $N=L^2=100$ Rydberg atoms in an optical tweezer array \cite{chen_continuous_2022}, and demonstrated true long-range FM order in 2D for the first time~\cite{chen_continuous_2022}. The effective Hamiltonian (with $J>0$) reads   
\begin{equation}\label{eq:Hxy+Hz}
H_{\rm XY} +\! H_{\rm Z}= -\frac{J}{2} \sum_{ (i,j)} \frac{1}{r_{{ij}}^3}\left(S_i^+ S_{j}^- + \rm{h.c.}\right) + \sum_{i} \delta_{i} S_{i}^z, 
\end{equation} 
where $S^\pm_{j}=S^x_{j}\pm i S^y_{j}$ refer to spin-$1/2$ operators encoded in two different Rydberg states with term symbols $60S_{1/2}$ and $60P_{1/2}$ for a $^{87}$Rb atom at site $\mathbf{r}_j$. The sum is over bonds $(i,j)$ and $r_{ij}=|\mathbf{r}_i-\mathbf{r}_j|$ \footnote{Note the different convention in defining the parameter $J$ in the experimental paper \cite{chen_continuous_2022}.}. The second term describes a staggered field $\delta_i = \pm \delta$ for site $i$ on the $A$ or $B$ sublattice, respectively. 
The initial state is designed to approximate a classical N{\'e}el state with spins pointing parallel or anti-parallel to the staggered field. Depending on the sign of $\delta$, this corresponds to a high- or low energy state of $H_\mathrm{Z}$. Subsequently, an exponential ramp reduces $|\delta|/J$ from its initial large value to zero and thus adiabatically prepares either low- or high energy states of the FM Hamiltonian, $H_\mathrm{XY}$. 
The latter can be interpreted as a low-energy state of the AFM Hamiltonian $-H_\mathrm{XY}$. 
Using this protocol, the experiment~\cite{chen_continuous_2022} explores the preparation of long-range ordered states on the FM side, and maps out a qualitative phase diagram for both the FM and the AFM. 

The ability to perform precise many-body thermometry on such an experiment, or more generally, any Rydberg systems with optical tweezers, has hitherto not been achieved.
Overcoming this challenge would provide key insights into the platform's ability to simulate more exotic and delicate phases of matter, such as elusive quantum liquids on frustrated lattices with AFM couplings, as well as enable more direct feedback and optimization of the experimental preparation fidelity.
Here we apply finite-temperature, equilibrium numerical methods to the $XY$ model with dipolar interactions.
From this we can estimate the initial and final entropies in experiment, and establish when the system thermalizes, with repercussions for future optical tweezer Rydberg experiments.

In the remainder of the paper we first discuss the ferromagnetic case in Sec.~\ref{sec:FM} and then turn to the antiferromagnetic side in Sec.~\ref{sec:AFM}. Conclusions are drawn in Sec.~\ref{sec:conclusion} and the appendices contain details on further comparisons to the experiment, high-temperature expansion, spin-wave theory and the computational methods.

\section{Ferromagnetic interactions $(J\!>\!0)$}
\label{sec:FM}
We employ quantum Monte Carlo (QMC) simulations with worm-type~\cite{Prokofev_worm_1998} updates to study the model, $H_{\rm XY} + H_{\rm Z}$ [see Eq.~\eqref{eq:Hxy+Hz}] on the square lattice of linear size $L$ with periodic boundary conditions.
The model is sign-free for the FM case, $J > 0$. The code is based on an adaptation of Ref.~\cite{Sadoune2022}, see App.~\ref{app:QMCsimulations} for details. 
There are two quantities convenient for studying spontaneous $U(1)$ symmetry breaking. The first one is the spin stiffness, $\chi_{\rm s}$, which in a path-integral formulation~\cite{Pollock1987} is related to the fluctuations of the winding number $\mathbf{W} = (W_x, W_y)$,
$\chi_{\rm s} = \left< W^2 \right> / (2 \beta)$.
Due to the long-range nature of the interactions, we work with odd values of $L$ in order to avoid ambiguities in $\mathbf{W}$ \footnote{Note that we are only interested in this quantity in the vicinity of a phase transition, more precisely, only in the fact that winding numbers are integers and therefore the scale-invariant quantity at the transition.}.

The second quantity is the in-plane magnetization squared \cite{chen_continuous_2022},
\begin{equation}
m_{\perp}^2 = \frac{1}{N} \sum_{j} C^{+-}( j),
\label{eq:def_condfrac}
\end{equation}
where $ C^{+-}( j) = \left< S^+_{j} S^-_{0} + {\rm h.c.} \right>$ is the equal-time off-diagonal spin correlation function for a system with translation invariance. 

\begin{figure}[t]
    \centering
    \includegraphics[width=\columnwidth]{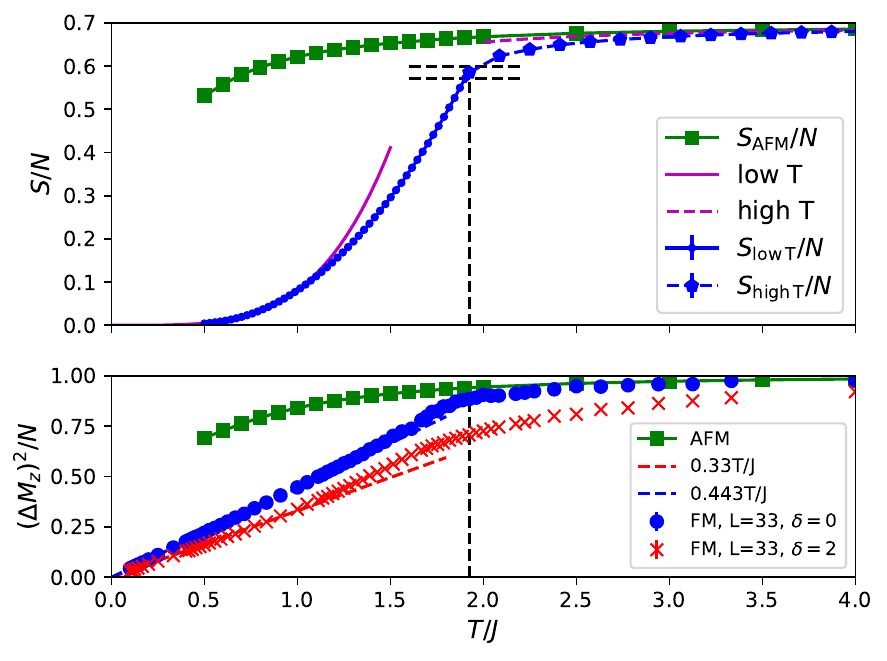}
    \caption{Upper: Entropy as a function of temperature for the dipolar system with $\delta=0$. The green curve for the AFM case is obtained from the free energy computed with pm-fRG; the blue curve is for the FM case where the entropy is determined from integration of energy (from QMC), Eq.~\eqref{eq:SfromE}. This is done separately in the disordered and ordered phases, each initiated by their analytically known asymptotic behavior (solid purple for spin waves and dashed purple for second order high temperature expansions, see App.~\ref{app:HTE}). The entropy per site at the FM critical temperature $T_c/J =1.923(1)$ (vertical dashed black line) is $S_c/N = 0.585(15)$ (horizontal dashed black line shows the uncertainty window). Lower: Variance of the z-magnetization as a function of temperature for the AFM case ($\delta = 0$) obtained by pm-fRG and for the FM case with $\delta =0$ and $\delta = 2$ computed by QMC for a system with linear system size $L=33$. The dashed lines depict linear-in-temperature behavior as predicted by spin-wave theory (see text).
    } 
     \label{fig:entropy}
\end{figure}

According to the theory of finite-size scaling for thermal transitions, the curves $\chi_{\rm s}(L)$ for the spin stiffness 
intersect at the critical temperature, up to scaling corrections. The same holds for the curves $m_{\perp}^2(L)L^{\eta}$ for the in-plane magnetization squared.  Whereas the critical exponents of long-range XY models decaying with exponents $\alpha \lesssim 4$ have long been debated~\cite{fisher_critical_1972,sak_recursion_1973}, the dipolar case exhibits, in line with the most recent review~\cite{defenu_long-range_2021},  $\nu=\eta=1$. Here, $\eta$ and $\nu$ are the anomalous- and correlation length critical exponents, respectively.

In Fig.~\ref{fig:QMC_scaling}(b) we show how the critical temperature for $\delta=0$ can be read off from the intersection points of the in-plane magnetization squared curves, rescaled with $L^{\eta = 1}$. Although we observe some corrections to scaling for small system sizes, the intersections for the larger system sizes $L$ fall onto a single point, within our resolution. We thus find the critical temperature $T_c/J=1.923(1)$. 
A finite-size analysis of the spin stiffness yields a compatible value; we refer to App.~\ref{app:QMCsimulations} for further information on the critical behavior. All our data are consistent with $\nu=\eta=1$.  We consequently repeated this type of analysis for several values of $\delta \in [0,4]$, from which we determine the phase diagram shown in Fig.~\ref{fig:QMC_scaling}(a).  

In order to analyze the $T=0$ quantum phase transition driven by $\delta$, we assume a dynamical exponent $z=1/2$ while $\nu=\eta=1$~\cite{defenu_long-range_2021} remain the same as before. This reduces the scaling forms to a one-parameter set of curves labeled by the linear system size $L$,  $\chi_s(L) \beta$ for the spin stiffness, and $m_{\perp}^2(L) L^{\eta}\beta$ for the in-plane magnetization squared. Each family of curves should intersect at the quantum critical point, $\delta_c$, for $L$ large enough. This is shown in Fig.~\ref{fig:QMC_scaling}(c).  
The corrections to scaling appear slightly larger than for the thermal phase transitions at small values of $\delta$, but the intersection points can still be extrapolated by a single powerlaw fit (not shown). From this we find the location of the quantum critical point at $\delta_c/J = 4.03(2)$, which is compatible with the rapid drop in $T_c$ shown in Fig.~\ref{fig:QMC_scaling}(a) setting in for $\delta \gtrsim 3$. The transition temperatures, as well as $\delta_c$ differ by a factor of two from the experimentally reported ones, mostly due to finite size effects (see below).

In addition to correlation functions, knowing the entropy as a function of temperature provides another key piece of information for characterizing the dipolar XY model. 
In the experimental, if preparation is assumed to be 
 adiabatic, then the entropy $S$ is  conserved.
 To this end, once one characterizes the entropy of the initial state, one can immediately  read off the  temperature and energy of the expected final state from the QMC data. 
 Knowledge of the entropy density at the critical temperature provides valuable information about the feasibility of experimentally reaching an ordered state.

However, entropies $S(T)$ are  cumbersome to compute in QMC simulations. One approach is to use a a fine grid of internal energy data, $E(T)$ and to compute \cite{Capogrosso2010}
\begin{equation}
    S(T_2) \!- \!S(T_1) =\! \frac{E(T_2) \!- \!E(T_1)}{T_2} +\! \int_{T_1}^{T_2} \!\frac{E(T') \!- \!E(T_1)}{T'^2} dT'.
    \label{eq:SfromE}
\end{equation}
%
In practice, we calculate $S(T)$ with both a low- and high-temperature starting point. For the low-temperature data, we apply linear spin wave theory (SWT) analysis: The energy is fitted by $E/N = E_0/N + \gamma T^5$ for small $T$ up to $T_{\rm low} = 0.5$, from which we extract $\gamma = 0.047(1)$ which is in reasonable agreement with the SWT prediction $\gamma^{SWT}\simeq 0.0393$ derived in App.~\ref{app:SWT}.
We also observe that finite-size effects strongly affect $E_0$ and $\gamma$.
Spin waves absorb very little entropy due their $S \sim T^4$ behavior. 
At high temperature, we calculated $E(T)$ to third order in $J/T$ in App.~\ref{app:HTE}. At intermediate temperatures we fitted a cubic spline through the $E/T$ and $E/T^2$ curves found from QMC, which was then integrated according to Eq.~\eqref{eq:SfromE}.
The difference in entropy at the critical temperature then gives a good measure for the accuracy of our approach.
We find the critical entropy per site $S(T_c)/N = 0.585(15)$, see Fig.~\ref{fig:entropy}. The entropy is seen to drop rapidly between $T=T_c$ and $T/J \simeq 1$.

\begin{figure}
    \centering
    \includegraphics[width=0.9\columnwidth]{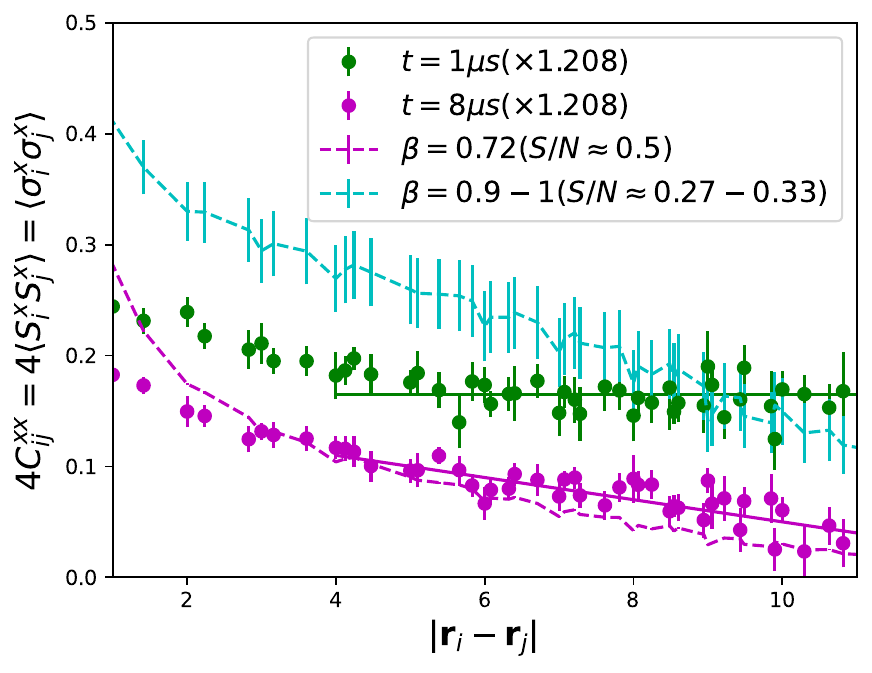}
    \caption{Experimental correlators for a $10\times 10$ system with open boundary conditions obtained for various durations $ t =1 \mu s$ (green dots) and $t = 8 \mu s$ (magenta dots), taken from Ref.~\cite{chen_continuous_2022} (and corrected for experimental uncertainties, see Ref.~\cite{chen_continuous_2022} and App.~\ref{app:comparisonToExp}).  QMC curves at temperatures corresponding to the entropy of the initial system (dashed cyan line  where error bars take the uncertainty on the initial state preparation into account) and the final system (dashed magenta line), obtained as a best possible fit. The green dots and green solid line (constant), seen in experiment for short durations $t \sim 1-2 \mu s$, are not compatible with the QMC, where the correlator decays linearly (cf.~magenta solid line).
    }
     \label{fig:QMC_corrfun_exptMAIN}
\end{figure}

Finally, building on our numerical results, we offer a number of new insights into the recent dipolar Rydberg array experiment, Ref.~\cite{chen_continuous_2022}.
We begin by computing the behavior of the spin correlation functions for $L=65$ with periodic
boundary conditions [Fig.~\ref{fig:correlator}(a)]; we note that these system sizes are significantly larger than those of the experiment. 
As the $4C_{ij}^{xx}=4\left\langle S_i^x S_j^x \right\rangle$ data in Fig.~\ref{fig:correlator}(a) indicate, a plateau appears for any $T<T_c$ and the sensitivity of the plateau value, $4C^{xx}_\infty$, which increases rapidly with lowering $T$ can serve as an excellent thermometer. For instance, the initial state in the experiment~\cite{chen_continuous_2022} with $N=42$ is $S_{\rm init}/N \approx 0.2$, which, via Fig.~\ref{fig:entropy} corresponds to the curve $T/J = 1.282$ in Fig.~\ref{fig:correlator}(a).

Such a plateau was seen in Ref.~\cite{chen_continuous_2022} and shown as the green line in Fig.~\ref{fig:QMC_corrfun_exptMAIN}, corresponding to data taken after $t=1 \mu$s. Contrary to expectations, it cannot be reproduced from an equilibrium simulation. The initial entropy of the experimental $10 \times 10$ system is $S/N = 0.30(3)$, and corresponds to the cyan spin correlation function in Fig.~\ref{fig:QMC_corrfun_exptMAIN}. It, surprisingly, has a linear slope, which is seen for any correlation function at higher temperature in the ordered phase. A plateau with such a low value $\sim 0.17$ as in experiment can not exist in equilibrium at this system size, and this observation must hence be a non-equilibrium effect. Experimental correlation functions for longer durations are reasonably well fitted by QMC results as shown by the magenta curves in Fig.~\ref{fig:QMC_corrfun_exptMAIN}. The linear slope is hence an equilibrium feature of this small system with open boundary conditions, defying textbook pictures of asymptotically constant correlation functions in the thermodynamic limit. The decay of the plateau at later times ($t > 1 \mu$s) was assumed in Ref.~\cite{chen_continuous_2022} to be driven by decoherence effects; here we advocate a picture of approaching (and returning after atom-light interactions, see below) to equilibrium at later times witnessed by the linearly decaying curves.

The experiment~\cite{chen_continuous_2022} also provides data for the variance of the z-magnetization, $(\Delta M_z)^2/N = ( \left< M_z^2 \right> - \left< M_z \right>^2)/N = 4 \sum_{i,j} \left\langle S_i^z S_j^z \right\rangle /N$. For $T/J\rightarrow \infty$ this approaches unity and for $T/J\rightarrow 0$ SWT predicts a linear $T-$dependence with a slope of $0.443$ (at $\delta = 0$), see App.~\ref{app:SWT}. The QMC results, shown in the lower panel of Fig.~\ref{fig:entropy}, reproduce both asymptotic behaviors. We thus suggest that $(\Delta M_z)^2/N$ can be used as a thermometer. \\
On the $L=10$ system with open boundary conditions, QMC likewise finds $(\Delta M_z)^2/N \propto T$ but with a steeper slope, see App.~\ref{app:comparisonToExp}. 
We note that since the Hamiltonian, Eq.~\eqref{eq:Hxy+Hz}, conserves the total excitation number,  there is no coupling between different $M_z$ sectors.
Although we believe that modeling the initial preparation via a grand-canonical ensemble is reasonable, dynamical processes could lead to an ensemble where the width of the $M_z$ distribution is not in equilibrium with the ensemble of states at a fixed magnetization. 
Nevertheless, the experimentally reported values for $(\Delta M_z)^2$ (FM, $N = 100$, late times) correspond to temperatures which are consistent  — within the $(\sim 20\%)$ spread of the experimental data — to the temperature extracted from the spin correlation functions.
This suggests the following picture for the non-equilibrium behavior observed in the experiment at intermediate times: Rare incoherent and $M_z$ non-conserving events broaden the $M_z$ distribution symmetrically and deposit a small amount of energy into the system, followed by each $M_z$ sector returning back to equilibrium under its own dynamics. The microscopic understanding of these events remains an open question.

\section{Antiferromagnetic interactions ($J < 0$)}
\label{sec:AFM}

In the AFM case, frustration due to long-range couplings prevents the use of QMC. Instead, we apply the recently developed pseudo-Majorana functional renormalization group (pm-fRG) \cite{niggemannFrustratedQuantum2021, niggemannQuantitativeFunctional2022, mullerPseudofermionFunctional2023}. This diagrammatic method is oblivious to frustration and builds on a faithful representation of spin-1/2 operators in terms of three Majorana fermions, $S_j^x=-i\eta_j^y\eta_j^z$, $S_j^y=-i\eta_j^z\eta_j^x$ and $S_j^z=-i\eta_j^x\eta_j^y$ \cite{tsvelikNewFermionic1992}. Spin Hamiltonians with bilinear spin-interactions are thus mapped to purely interacting Majorana systems which are then treated by the functional renormalization group \cite{metznerFunctionalRenormalization2012} with a Matsubara frequency cutoff. So far, pm-fRG has only been applied to Heisenberg systems with full $SO(3)$ spin rotation symmetry, and the modifications required to treat the $U(1)$ symmetric case as well as further details on pm-fRG are given in App.~\ref{app:pm-fRG}. As the inclusion of magnetic fields is  more complicated we here focus on the case $\delta=0$.

\begin{figure}[t]
    \centering
    \includegraphics[width=\columnwidth]{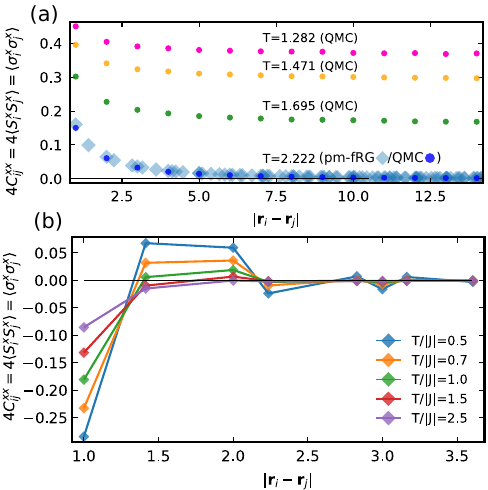}
    \caption{Equilibrium equal-time correlator $C_{ij}^{xx}$ versus distance. (a) FM case: The data are computed with QMC (dots) for linear system size $L=65$ and for various temperatures above and below the critical temperature $T_c/J=1.923(1)$. Statistical error bars are smaller than the symbol size. 
    The magenta curve corresponds to an entropy comparable to the initial state entropy in the (much smaller) $N=42$ experimental system, the green curve is close to the one reported in Fig.~3(a) of Ref.~\cite{chen_continuous_2022} after readout errors have been taken into account (this agreement is coincidental, see text).
    The pm-fRG data (diamonds) for $T/J=2.222$ are shown for comparison. (b) AFM case: The data are computed with pm-fRG for an essentially infinite system. Lines are guides to the eye. }
     \label{fig:correlator}
\end{figure}
As (pm-)fRG involves a truncation of the hierarchy of flow equations, it is important to carefully gauge the validity of the results as $T/|J|$ is lowered towards the strong-coupling regime. As a first check of pm-fRG, we resort to the FM case and compare $C_{ij}^{xx}$ from pm-fRG to the error controlled QMC results. The agreement is excellent for $T/J=5$ (data not shown) while differences of just a few percent are found close to the phase transition at $T/J=2.222$, see Fig.~\ref{fig:correlator}(a). While entering an ordered phase with fRG is difficult, the critical temperature can easily be found \cite{niggemannQuantitativeFunctional2022} and we obtain $T_c^\text{pm-fRG}/J=2.02$, only $5\%$ off from the QMC result and with matching critical exponents.

On the AFM side, we resort to two independent internal consistency checks between 2- and 4-point vertices detailed in App.~\ref{app:pm-fRG}. First, we compare $E(T)$ computed from the separate flow equation of the free energy \cite{niggemannFrustratedQuantum2021} and via $\left\langle H_\mathrm{XY} \right\rangle$ using the equal-time correlators. The second check invokes an exact constant of motion related to the pm representation. Comparing the outcome of these consistency checks to other models where exact benchmark data are available (FM-side, small spin clusters) suggests that pm-fRG data are reliable with a few-percent error for $T/|J| \ge 0.5$. The results for the correlator $C^{xx}_{ij}$ at various temperatures between $T=2.5|J|$ and $T=0.5|J|$ are reported in Fig.~\ref{fig:correlator}(b). Note that the full checkerboard-like AFM correlation pattern only builds up at low enough $T$; for example, the next-nearest-neighbors become positively correlated only below $T\simeq 1.0|J|$. 
The entropy $S(T)$ is straightforwardly obtained from the free energy mentioned above, see Fig.~\ref{fig:entropy}(a). We also provide the pm-fRG estimate for $(\Delta M^z)^2(T)$, see Fig.~\ref{fig:entropy}(b). 
In App.~\ref{app:comparisonToExp}, we fit the short-distance parts of the experimentally measured correlators to thermal pm-fRG data.

\section{Conclusion}
\label{sec:conclusion}

Using large-scale numerical methods we compute the thermodynamics for dipolar spin-$1/2$ models with $XY$ symmetry and staggered field $\delta$, with emphasis on the experimentally measurable spin correlation functions and the variance of the z-magnetization. We also revealed the $T-\delta$ phase diagram on the FM side where the critical temperature $T_c = 1.923(1)J$ at $\delta=0$. The corresponding entropy per particle is $S_c/N = 0.585(15)$. The zero-temperature FM transition is found at $\delta_c/J = 4.03(2)$.  
Our numerical simulations confirm that the entropy density of the resulting thermal state (on the FM side) at late times in the experiment of  Ref.~\cite{chen_continuous_2022} places it in the ordered phase in the thermodynamic limit.  
However, our results also suggest that the experimentally observed correlation function plateau could be an intrinsically non-equilibrium feature. 
Understanding the origin of this intermediate time plateau, and whether such a plateau generally emerges as a non-equilibrium pre-cursor to order, is an intriguing open question.
More broadly, characterizing the dynamical adiabatic preparation of many-body phases and understanding the features that govern the approach to equilibrium represents an overarching theme that will be crucial for pushing the frontiers of Rydberg-based quantum simulations, and quantum spin liquids in particular~\cite{giudici2022dynamical,sahay2022quantum}.


\begin{acknowledgements}
We thank I.~Bloch, G.~Bornet, A.~Browaeys, C.~Chen, N.~Defenu, G.~Emperauger, C.~Groß, T.~Lahaye, A.~Läuchli, V.~Liu, M.~Zaletel for useful discussions. We thank A.~Browaeys and T.~Lahaye for allowing us to reproduce unpublished experimental results in App.~\ref{app:comparisonToExp}. 
Numerical data for this paper are available under \url{https://github.com/LodePollet/QSIMCORR}.
The QMC code makes use of the ALPSCore libraries~\cite{gaenko_updated_2017,wallerberger_updated_2018}.
We acknowledge funding by the Deutsche Forschungsgemeinschaft (DFG, German Research Foundation) under Germany's Excellence Strategy -- EXC-2111 -- 390814868. LP is supported by FP7/ERC Consolidator Grant  No.~771891 (QSIMCORR).
BS gratefully acknowledges the Gauss Centre for Supercomputing e.V.~(www.gauss-centre.eu) for funding this project by providing computing time through the John von Neumann Institute for Computing (NIC) on the GCS Supercomputer JUWELS at Jülich Supercomputing Centre (JSC). B.S.~is supported by a MCQST-START fellowship and by
the Munich Quantum Valley, which is supported by the
Bavarian state government with funds from the Hightech
Agenda Bayern Plus.
N.Y.Y.~acknowledges support from the Air Force Office of Scientific Research through the MURI program (FA9550-21-1-0069). 
M.B.~and S.C.~acknowledge support from the U.S.~Department of Energy, Office of Science, National Quantum Information Science Research Centers, Quantum Systems Accelerator (QSA).

\end{acknowledgements}

\appendix
\section{Comparison to experiment}
\label{app:comparisonToExp}

\subsection{FM case}

In this section we re-analyze the spin correlation function obtained in experiment in the protocol~\cite{chen_continuous_2022} expected to be adiabatic. The experimental system consisted of a $10 \times 10$ square lattice with open boundary conditions. We note that the $6 \times 7$ system was already analyzed in great detail in Ref.~\cite{chen_continuous_2022}. The system was prepared using optical tweezers in a N{\'e}el state in a strong staggered field with $2\delta / J = 9 {\rm MHz} / 0.77 {\rm MHz} = 11.69$ (note again the different unit convention resulting in a factor 2 difference with Ref.~\cite{chen_continuous_2022}). 
The staggered field was ramped down exponentially with a time constant $\tau = 0.3 \mu$s. 
The spin correlation function was plotted as a function of radial distance for several durations of the experiment. 
The strongest correlations were observed at $t = 1 \mu s$. These notably show a plateau for distances beyond 5 lattice spacings, indicative of the finite-size correlation function of a symmetry-broken state.
For larger durations, the plateau disappears and one observes a linear decay of correlations with distance. At $t=8 \mu s$, the signal has dropped by about 25\% compared to $t = 1 \mu s$. \\

\begin{figure}
    \centering
    \includegraphics[width=\columnwidth]{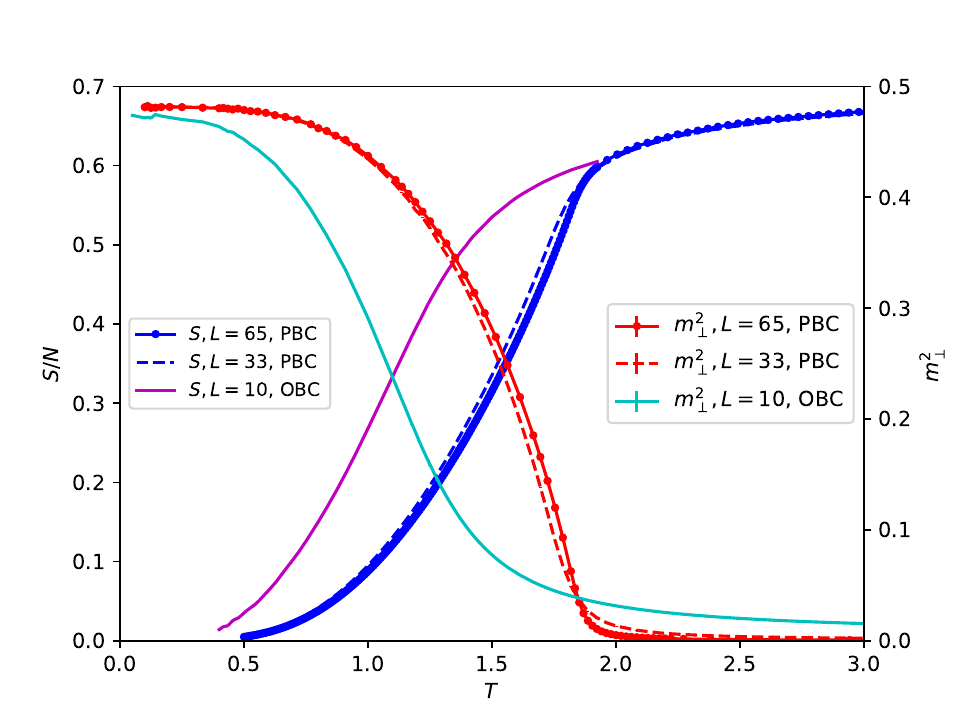}
    \caption{Entropy per site (left axis) and $m^2_{\perp}$ (right axis) for systems of size $L=33$ and $L=65$ with periodic boundary conditions (PBC) and for a system of size $L=10$ with open boundary conditions (OBC). 
    }
     \label{fig:QMC_FSS}
\end{figure}

The entropy per particle of the distribution of the initially prepared N{\'e}el states  is estimated as $S_{\rm init}^{N=100}/N = 0.30(3)$ for $N=100$ and $S^{N=42}_{\rm init}/N \approx 0.2$ for $N=42$. These estimates are justified as follows. Restricting ourselves to the A-sublattice and neglecting holes, for $N=42$ this is the configurational entropy associated with the measured magnetization, $\left< Z_A \right> = 0.89$. For $N=100$,  no value is provided in Ref.~\cite{chen_continuous_2022}, but taking the higher values of $\eta_A=0.07(1)$ and $\eta_B=0.10(1)$ into account (cf. Table I in Ref.~\cite{chen_continuous_2022} for their meaning), which the authors of Ref.~\cite{chen_continuous_2022} communicated to us, the  magnetization is considerably lower, $\left< Z_A \right> \sim 0.81(1)$, and the resulting configurational entropy considerably higher than for $N=42$. \\

\begin{figure}
    \centering
    \includegraphics[width=\columnwidth]{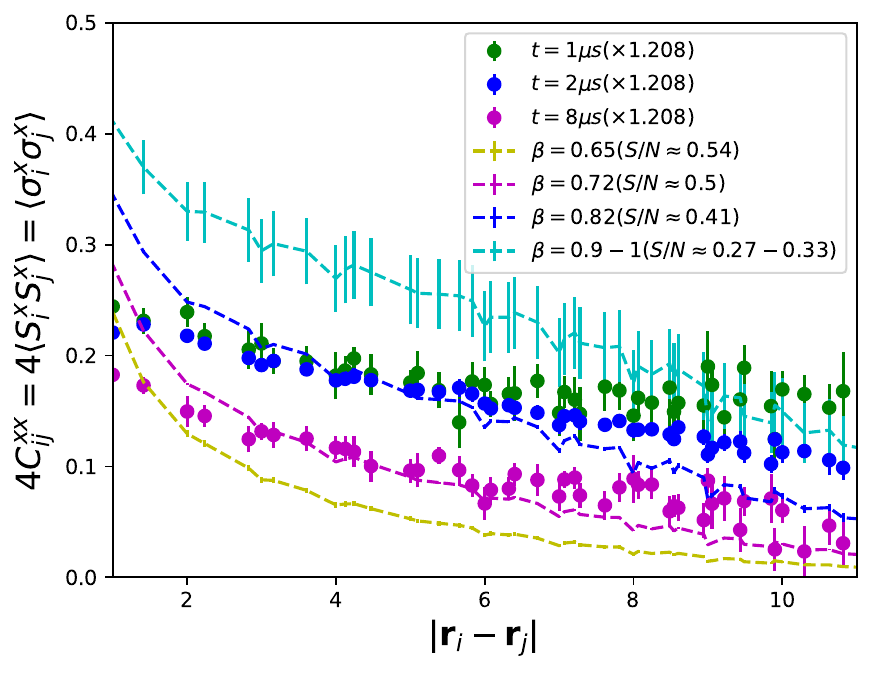}
    \caption{Dots denote spin correlation functions obtained for various experimental durations $ t =1 \mu s$ (green), $t = 2 \mu s$ (blue) and $t = 8 \mu s$, taken from Ref.~\cite{chen_continuous_2022} (and multiplied by 1.208, see text and Ref.~\cite{chen_continuous_2022}). Note that the values of the experimental data at short distances are underestimated because of the presence of holes and the fact that $H_{\rm XY}$ is active during the $X$ to $Z$ rotation. Also shown are the spin correlation functions obtained in quantum Monte Carlo simulations (dashed lines) for a $10 \times 10$ system with open boundary conditions, zero magnetic field, and zero staggered field $\delta$. 
     The radial averaging is performed over all sites of the lattice. We argue that an equilibrium approach is justifiable for times  $t \gtrsim 2 \mu$s. The cyan data indicate the range of values compatible with the error bars on the initial entropy, $S/N = 0.30(3)$. The dashed yellow curve is incompatible with the experimental data, and therefore provides a clear upper bound to the temperature in experiment. 
    }
     \label{fig:QMC_corrfun_expt}
\end{figure}

\begin{figure}
    \centering
    \includegraphics[width=\columnwidth]{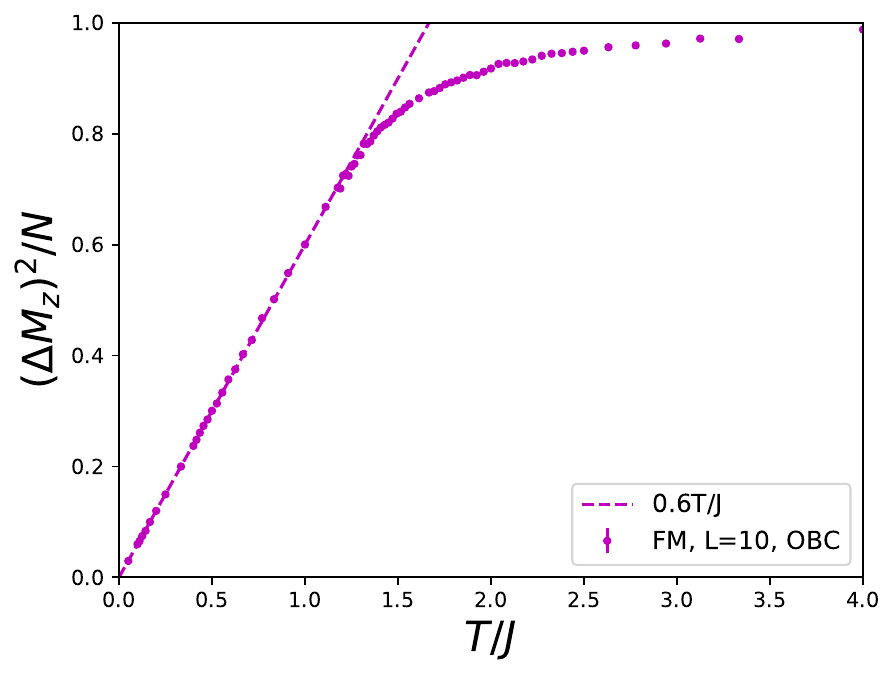}
    \caption{Variance of the z-magnetization per site as a function of temperature for a system of linear size $L=10$ with open boundary conditions inspired by the experimental setting~\cite{chen_continuous_2022}. The linear temperature dependence, predicted by linear spin-wave theory, is respected at low temperatures but with a different slope because of the finite system size. The linear temperature dependence holds however far beyond the regime of linear spin-wave theory (almost everywhere in the ordered phase), and thus enables experimental thermometry independent of numerical simulations, provided that the data are in equilibrium.
    }
     \label{fig:QMC_variance_OBC}
\end{figure}
We simulated the experimental $L=10$ system in equilibrium with QMC simulations and show the entropy and $m^2_{\perp}$ in Fig.~\ref{fig:QMC_FSS} where we also compare to $L=33$ and $L=65$. 
It is apparent that the finite size effects are substantial for the experimental system size. We checked that excluding the boundary sites in the radial averaging in order to obtain the spin correlation functions (and thus $m^2_{\perp}$) increases its value by a factor $\sim 1.1$ (as compared to Fig.~\ref{fig:QMC_FSS} where boundary sites are included), but does not change the shape of the function and is therefore not a dominant factor. 
A direct comparison with experimental data is however complicated due to a number of factors; most notably (i) the detection errors (up to $10\%$) lead to a correction of the value of the correlations that amounts to up to $20\%$~\cite{chen_continuous_2022}, which is the same for $N=42$ and $N=100$ systems,
(ii) the measurement of correlations in the $x$-direction requires a rotation, but this is done in the presence of $H_{\rm XY}$, which generates additional interaction terms that are expected to have a $r^{-3}$ dependence, (iii) one expects $5$ \% holes in the system due to failed Rydberg excitations, which affect both initial entropies and read-out measurements, (iv) the preparation protocol involves ramps that are expected to be close to be adiabatic, (v) the many-body effects of spontaneous emission, jitter, etc are not quantified, and lie outside the model Hamiltonian.\\

An attempted comparison of the spin correlation functions between experiment and QMC simulations is shown in Fig.~\ref{fig:QMC_corrfun_expt}. This can not be seen as a full benchmark however, since we did not take the holes into account, and the measurement errors are not fully understood due to non-commuting terms. We further multiplied the experimental data from Ref.~\cite{chen_continuous_2022} by a factor $1.208$, which is the read-out error~\cite{chen_continuous_2022}. But we expect that the leading behavior can be caught, for not too small distances at least. As we will see below, the results are highly counter-intuitive because of the too small system size.  The most striking observation is that none of the relevant QMC correlation functions show a plateau, not even at temperatures as low as the ones corresponding to $S_{\rm init}/N$: The system size is too small to observe a plateau in equilibrium at those temperatures! We therefore conclude that the experimental data for $t = 1 \mu s$ are not in equilibrium. 

In fact, the shape of the experimental data at later times $t \gtrsim 2 \mu s$ is much closer following the QMC calculations. In Fig.~\ref{fig:QMC_corrfun_expt} we attempted to find the temperatures that reasonably match the experimental data for $t = 2 \mu s$ and $t = 8 \mu s$ (and in between, not shown). The QMC data are therefore suggestive of an alternative picture for the experimental data. We believe that at $t \gtrsim 2 \mu s$ the system is close to equilibrium, and that a thermal ensemble is a reasonable approximation to describe the system for larger times, at least up to $t = 8 \mu s$. We also find in the QMC (see Fig.~\ref{fig:QMC_variance_OBC}) that $ (\Delta M_z)^2$ takes $73(3) \%$ of its infinite temperature value, in agreement with the experimental result at $t \ge 2 \mu s$~\cite{chen_continuous_2022}, but not for $t = 1 \mu s$, where it is much closer to the $t=0 \mu s$ value, $\sim 60$\% (and the latter also agrees with our $S_{\rm init}$). The value for the variance computed by QMC relevant to the $t = 8 \mu s$ duration, is $0.81$.  
Based on the experimental information, we expect one spontaneous emission event resulting in a loss on a $10 \times 10$ lattice every $\mu s$, creating a defect of energy scale $\sim J$. 
The value of $J/2 = 0.77 $ MHz is indicative of a  strong tendency to quasi-equilibrium at increasing temperatures, cf.~the discussions in Ref.~\cite{Pichler2010,Schachenmayer2014}.
Other sources of decoherence (some of which are non-extensive) seem to be at work at similar frequencies. It is therefore conceivable that phase coherence builds up in the first $\mu s$, remains incomplete and is largely interrupted at the time scale $\sim 1 \mu s$. If we can approximate the experimental system (cf.~Fig~\ref{fig:QMC_corrfun_expt}) for $t = 2 \mu s$ with a thermal ensemble at $\beta = 0.82$ $(S/N = 0.41(2))$  and the one for $t = 8 \mu s$ with a thermal ensemble at $\beta = 0.72 $ $(S/N = 0.50(2))$, then we can estimate the entropy production as $\dot{S}/N \approx 0.015(3)$ per $\mu s$, which is rather low. If we separate ramp dynamics from such heating sources during the first $2 \mu$s (and take the latter again as constant), then the contribution to the change in entropy from the ramps is just $0.41(2) - 0.30(3) - 0.03(1) = 0.08(6)$. Although this is not entirely negligible, it is comparable to other sources of decoherence and certainly small compared to the entropy originating from the initial state preparation. This is consistent with the analysis of Ref.~\cite{chen_continuous_2022}, which numerically found that the spin-neutral energy gaps for this protocol were large (i.e.~of order $J$). \\

There are two big questions that will be left for future work. The first one is how the system thermalizes, and under what conditions a grand-canonical approach is justified. 
The second one is what might explain the experimental out-of-equilibrium correlation function at $t = 1 \mu s$.  Whether the experimentally observed plateau can be given any dynamical meaning is certainly an interesting question.


\subsection{AFM case}
\begin{figure}
    \centering
    \includegraphics{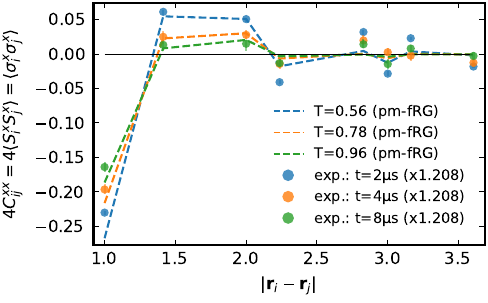}
    \caption{AFM case: Experimental spin correlation functions measured after $t =2 \mu s$ (blue dots), $t = 4 \mu s$ (orange dots) and $t = 8 \mu s$ (green dots), obtained from the authors of Ref.~\cite{chen_continuous_2022} (private communication). As in the FM case, we have multiplied the measured data with a factor of 1.208 to take into account measurement errors. The data for small distances are well reproduced by thermal pm-fRG simulations (dashed line, infinite system) at temperatures $T/|J|=0.56$, $0.78$ and $0.96$, respectively. The corresponding entropies are $S/N=0.550$, $0.596$ and $0.618$. 
    }
     \label{fig:pm-fRG_XX_Dipolar_AFM_Cxx_PAPER}
\end{figure}

We now turn to the AFM case where the correlations in the experimental $L=10$ system were found to be short range. We thus expect that finite-size corrections are much less important than on the FM side and compare the experimental data to the pm-fRG results obtained for an essentially infinite and translation invariant system. While pm-fRG can also handle small finite systems with open boundaries \cite{niggemannFrustratedQuantum2021}, treating an $L=10$ system (without translation symmetry) is currently beyond the method's capabilities.

The thermometry of the experimental results at $t=2 \mu s$, $t=4 \mu s$ and $t= 8\mu s$ (courtesy of the authors of Ref.~\cite{chen_continuous_2022}) is shown in Fig.~\ref{fig:pm-fRG_XX_Dipolar_AFM_Cxx_PAPER}. We match the short-range correlations for $|\mathbf{r}_i - \mathbf{r}_j|\leq 2$ to pm-fRG data obtained at $T/|J|=0.56$, $0.78$ and $0.96$, respectively. For distances beyond two lattice spacings the experimental signal is very weak but still significantly larger than the pm-fRG results. Comparing the entropies of the temperatures associated to $t=2 \mu s$ and $t= 8\mu s$ (c.f.~Fig.~2 of the main text) we obtain a heating rate of $\dot{S}/N = 0.0113$ per $\mu s$, in good agreement with the FM case. The experimental value of $(\Delta M^z)^2/N = 0.57(2)$ at $t=2 \mu s $ \cite{chen_continuous_2022} is about $20 \%$ smaller than the value $0.71$ obtained from the pm-fRG at $T=0.56$.

A possible explanation for this discrepancy could be additional experimental imperfections (like imperfect rotation for the measurement in the X-basis due to the presence of $H_\mathrm{XY}$, see the discussion above) that apparently lead to an underestimation of short-range correlations. This would mean that the temperatures extracted in Fig.~\ref{fig:pm-fRG_XX_Dipolar_AFM_Cxx_PAPER} are be underestimated, and the true correlations including data beyond two lattice spacings could be fitted by smaller temperatures. It is plausible that this temperature, which is currently out of reach for pm-fRG would also be consistent with the measured value of $(\Delta M^z)^2/N = 0.57(2)$.

\section{High temperature expansion}
\label{app:HTE}

The high-temperature expansion (HTE) of the energy $E(T)$ serves to check our numerical method implementation and is used to facilitate the entropy calculation from the energy following Eq.~(4) of the main text. The results are valid for both signs of $J=\pm|J|$ (FM/AFM) but we restrict ourselves to $\delta = 0$.

\begin{figure}
    \centering
    \includegraphics[width=\columnwidth]{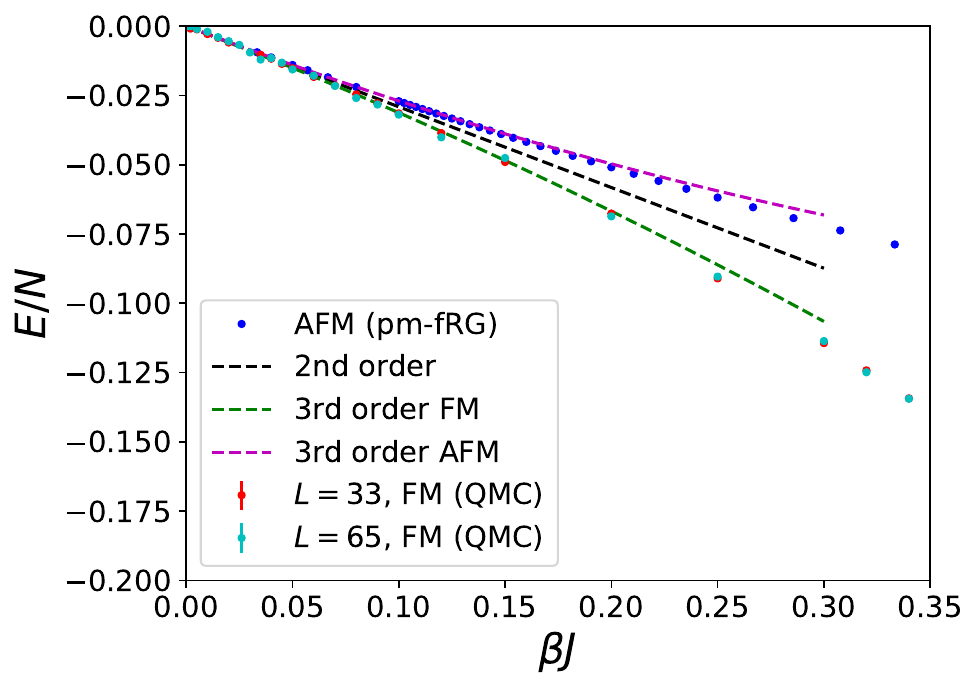}
    \caption{Energy per site as a function of inverse temperature $\beta J$ in the high temperature regime. Results are shown for the AFM case obtained by pm-fRG (blue dots) and for the FM case obtained by QMC (red and cyan dots for $L=33$ and $L=65$, respectively, with periodic boundary conditions and using the bare potential). The energies are compared with the predictions of Eq.~\eqref{eq:HTE} with truncations to second order (dashed black line) and third order (dashed green line for the FM case and dashed purple line for the AFM case).}
     \label{fig:QMC_energy_highT}
\end{figure}

The energy is expressed as $E=\mathrm{tr}\left[He^{-\beta H}\right]/\mathrm{tr}\left[e^{-\beta H}\right]$
which we expand to order $\beta^{2}$ using $\mathrm{tr}\left[H\right]=0$,
\begin{equation}
E=-\beta\frac{\mathrm{tr}\left[H^{2}\right]}{\mathrm{tr}\left[1\right]} + \frac{1}{2}\beta^{2}\frac{\mathrm{tr}\left[H^{3}\right]}{\mathrm{tr}\left[1\right]}+\mathcal{O}(\beta^{3}).
\end{equation}
We find 
\begin{equation}
\mathrm{tr}\left[H^{2}\right]=\frac{N}{2}2^{N-2}2\left(\frac{J}{2}\right)^{2}\sum_{\mathbf{r}\neq0}\left(\frac{1}{r^{3}}\right)^{2},
\end{equation}
where we considered clusters of two (different) sites. The first term
avoids overcounting of the bonds, the second takes into account the
trace over sites outside of the cluster. The third factor captures
the fact that the two cluster-basis states $\left|\uparrow,\downarrow\right\rangle $
and $\left|\downarrow,\uparrow\right\rangle $ contribute to the trace
and the remaining factors come from the Hamiltonian and the cluster
sum $\sum_{\mathbf{r}\neq0}1/r^{6} \equiv \phi_{2}\simeq4.6589$. Likewise,
\begin{equation}
\mathrm{tr}\left[H^{3}\right]\!=\!\frac{N}{3}2^{N-3}6\left(\frac{J}{2}\right)^{3}\!2 \, \frac{1}{2}\,\!\underset{\equiv\phi_{3}\simeq13.6527}{\underbrace{\sum_{\mathbf{r}_{1,2}\neq0,\mathbf{r}_{1}\neq\mathbf{r}_{2}} \!\!\! r_{1}^{-3}r_{2}^{-3}|\mathbf{r}_{1}-\mathbf{r}_{2}|^{-3}}}
\end{equation}
where we now have $6$ contributing states $\left|\uparrow,\uparrow,\downarrow\right\rangle ,\left|\uparrow,\downarrow,\uparrow\right\rangle ,\left|\downarrow,\uparrow,\uparrow\right\rangle $
(and all spins flipped) to the trace of the 3-site cluster. The rightmost
factor of $2$ indicates the two possible pathways of flipping a state
back to itself and $1/2$ avoids overcounting (due to $\mathbf{r}_{1}\leftrightarrow\mathbf{r}_{2}$).
In summary, with $\mathrm{tr}[1]=2^N$, we obtain (see Fig.~\ref{fig:QMC_energy_highT})
\begin{equation}
E/N=-\frac{\beta J^{2}}{16}\phi_{2} + \frac{\beta^{2}J^{3}}{64}\phi_{3}+\mathcal{O}(\beta^{3}). \label{eq:HTE}
\end{equation}
From Eq.~(4) of the main text, we conclude for the entropy
\begin{equation}
S/N=\frac{J^2 (J \phi_3-3 T \phi_2)}{96 T^3}+\log (2). \label{eq:HTE-S}
\end{equation}
In order not to overburden the entropy plot Fig.~2 in the main text due to the different signs for the FM and AFM case in third order, we only plotted the HTE to second order. In the actual calculations leading to Fig.~2 in the main text we used the more precise third order results.

\section{Spin wave theory (FM case)}
\label{app:SWT}

In this section we detail the spin-wave theory (SWT) calculations for the low-temperature behavior of the internal energy 
\begin{equation}
    E(T) = \langle H \rangle \simeq E(T=0)+\gamma \frac{T^5}{|J|^4}
\end{equation}
and the fluctuations of the z-magnetization ($M^z=2\sum_i S_i^z$ \cite{chen_continuous_2022}),
\begin{equation}
    (\Delta M^z)^2 /N = 4 \sum_i \langle S_i^z S_j^z \rangle \simeq \kappa\frac{T}{|J|} .
\end{equation}
The SWT for dipolar XY models has first been considered in Ref.~\cite{Peter2012}, it builds on the representation of spin-1/2 operators in terms of Holstein-Primakoff bosons, $S_{i}^{x}=-1/2+a_{i}^{\dagger}a_{i}$, $S_{i}^{y}\simeq(a_{i}-a_{i}^{\dagger})/(2i)$, $S_{i}^{z}\simeq(a_{i}+a_{i}^{\dagger})/2$ and the Fourier transform $a_{i}=\frac{1}{\sqrt{N}}\sum_{\mathbf{q}}e^{i\mathbf{q}\cdot\mathbf{r}_{i}}a_{\mathbf{q}}$. The non-interacting part of the spin-wave Hamiltonian reads \cite{Peter2012}
\begin{eqnarray}
        H	&=&	E_{0,MF}+\frac{1}{2}\sum_{\mathbf{q}}E_{\mathbf{q}}+\sum_{\mathbf{k}}E_{\mathbf{q}}b_{\mathbf{q}}^{\dagger}b_{\mathbf{q}}, \\
        E_{\mathbf{q}}	&=&	\frac{1}{2}|J|\sqrt{\varepsilon_{\mathbf{0}}\left(\varepsilon_{\mathbf{0}}-\varepsilon_{\mathbf{q}}\right)}\simeq\frac{1}{2}|J|\sqrt{2\pi q|\varepsilon_{\mathbf{0}}|},
\end{eqnarray}
where $E_{0,MF}=-|J|\frac{\varepsilon_{0}}{8}N$ is the classical ground state energy and $\varepsilon_{\mathbf{0}}\simeq9.033$ and the precise form of $\varepsilon_{\mathbf{q}}$ is given in Ref.~\cite{Peter2012}. The bosonic operators $b_\mathbf{q}$ are defined via a Bogoliubov transformation, $a_{\mathbf{q}}=u_{\mathbf{q}}b_{\mathbf{q}}+v_{\mathbf{q}}b_{-\mathbf{q}}^{\dagger}$ and $a_{-\mathbf{q}}^{\dagger}=v_{\mathbf{q}}b_{\mathbf{q}}+u_{\mathbf{q}}b_{-\mathbf{q}}^{\dagger}$ where $u_\mathbf{q}=u_\mathbf{-q}=\mathrm{cosh \theta_\mathbf{q}}$, $v_\mathbf{q}=v_\mathbf{-q}=\mathrm{sinh \theta_\mathbf{q}}$ and $\tanh(2\theta_{\mathbf{q}})=-\frac{\varepsilon_{\mathbf{q}}}{2\varepsilon_{\mathbf{0}}-\varepsilon_{\mathbf{q}}}\simeq-\frac{\varepsilon_{0}-2\pi q}{\varepsilon_{0}+2\pi q}$.\\

To find the spin wave contribution to the energy per spin, we compute
\begin{equation}
\frac{E(T)-E(T=0)}{N}=\frac{1}{N}\sum_{\mathbf{q}}\frac{E_{\mathbf{q}}}{e^{\beta E_{\mathbf{q}}}-1}
\end{equation}
We consider small $T$, expand for small $q$, take $N\rightarrow\infty$ and extend the Brillouin zone integration to infinity,
\begin{widetext}
\begin{equation}
\frac{E(T)-E(T=0)}{N}\simeq\frac{1}{2\pi}\int_{0}^{\infty}q dq\,\frac{\frac{1}{2}|J|\sqrt{2\pi q|\varepsilon_{\mathbf{0}}|}}{e^{\beta\frac{1}{2}|J|\sqrt{2\pi q|\varepsilon_{\mathbf{0}}|}}-1}=\frac{T}{\pi}\left[\frac{2T^{2}}{|J|^{2}\pi|\varepsilon_{\mathbf{0}}|}\right]^2\int_{0}^{\infty}dx\,\frac{x^{4}}{e^{x}-1}=\underset{\equiv\gamma}{\underbrace{\frac{96\zeta(5)}{\pi^{3}\varepsilon_{\mathbf{0}}^{2}}}}\frac{T^{5}}{|J|^{4}},
\end{equation}
and read off $\gamma\simeq0.03935$.

For the fluctuations of the z-magnetization, we use
\begin{equation}
    \left\langle S_{i}^{z}S_{j}^{z}\right\rangle =\frac{1}{4}\left\langle (a_{i}+a_{i}^{\dagger})(a_{j}+a_{j}^{\dagger})\right\rangle = \frac{1}{4N}\sum_{\mathbf{q}}e^{i\mathbf{q}\cdot(\mathbf{r}_{i}-\mathbf{r}_{j})}\left(1+2\left\langle b_{\mathbf{q}}^{\dagger}b_{\mathbf{q}}\right\rangle \right)\left(v_{\mathbf{q}}+u_{\mathbf{q}}\right)^{2}
\end{equation}
and substitute $\left(v_{\mathbf{q}}+u_{\mathbf{q}}\right)^{2}\simeq\sqrt{\frac{2\pi}{\varepsilon_{0}}q}$. For $T=0$, we have
\begin{eqnarray}
\left(\Delta M_{z}(T=0)\right)^{2}/N & = & 4\sum_{i}\left\langle S_{i}^{z}S_{j}^{z}\right\rangle _{T=0}=\sqrt{\frac{2\pi}{\varepsilon_{0}}}\sum_{\mathbf{q}}\sqrt{q}\frac{1}{N}\sum_{i}e^{i\mathbf{q}\cdot(\mathbf{r}_{i}-\mathbf{r}_{j})}=0,
\end{eqnarray}
while for $T>0$, the asymptotic $\sqrt{q}$ in the last equation is replaced by a
constant. We take $\beta<\infty$ large but fixed
and obtain
\begin{eqnarray*}
\left(\Delta M_{z}\right)^{2}/N & = & 4\sum_{i}\left(\left\langle S_{i}^{z}S_{j}^{z}\right\rangle _{T}-\left\langle S_{i}^{z}S_{j}^{z}\right\rangle _{T=0}\right)\\
 & = & \sqrt{\frac{2\pi}{\varepsilon_{\mathbf{0}}}}\frac{1}{N}\sum_{i}\sum_{\mathbf{q}}\frac{2\sqrt{q}}{e^{\beta E_{q}}-1}e^{i\mathbf{q}\cdot(\mathbf{r}_{i}-\mathbf{r}_{j})}\\
 & \simeq & T\sqrt{\frac{2\pi}{\varepsilon_{\mathbf{0}}}}\sum_{\mathbf{q}}\frac{2\sqrt{q}}{\frac{1}{2}|J|\sqrt{2\pi q\varepsilon_{\mathbf{0}}}}\frac{1}{N}\sum_{i}e^{i\mathbf{q}\cdot(\mathbf{r}_{i}-\mathbf{r}_{j})}\\
 & = & \underset{\kappa}{\underbrace{4/\varepsilon_{\mathbf{0}}}}\frac{T}{|J|}
\end{eqnarray*}
where $\kappa\simeq0.443$.

\section{Quantum Monte Carlo simulations (FM case)}
\label{app:QMCsimulations}

\subsection{Algorithm}
Path Integral quantum Monte Carlo (QMC) methods are among the most successful numerical methods to study (effectively) bosonic systems of finite extent in thermodynamic equilibrium with $T = 1/\beta > 0$.
To set up the theoretical framework it is convenient to write the Hamiltonian as $H = H_0 - H_1$,
where $H_0 = H_{\rm Z}$ is diagonal in the computational basis and $H_1 = -H_{\rm XY}$ causes a transition from one basis state to another. We work in the computational basis, defined as the eigenstates of the spin-1/2 $S^z$ operators on every site.
The algorithm is based on  the perturbative time-ordered expansion of the partition function in continuous imaginary time,
\begin{equation}
Z = {\rm Tr}\ e^{- \beta H} = {\rm Tr}\ e^{- \beta H_0} \sum_{n=0}^{\infty} \int_0^{\beta} d\tau_1 \int_0^{\tau_1} d\tau_2 \ldots \int_0^{\tau_{n-1}} d\tau_n \,\, H_1(\tau_1) \ldots H_1(\tau_n).
\label{eq:Z_pert}
\end{equation}
\end{widetext}
The trace is taken with respect to all states in the computational basis. 
The positivity of this expansion is necessary and requires $J > 0$ ({\it i.e.}, the dipolar spin-exchange terms must be ferromagnetic).

The Heisenberg operators are defined as
\begin{equation}
H_1(\tau_k) = e^{\tau_k H_0} H_1 e^{-\tau_k H_0}.
\label{eq:Heisenberg}
\end{equation}
The central quantity of interest in the worm algorithm~\cite{Prokofev_worm_1998} is the single-particle Green's function $G(A, \tau_A ; B, \tau_B)$ defined as the thermal average $G(A, \tau_A ; B, \tau_B) = \frac{1}{Z} \left< \mathcal{T} [ S^+_A(\tau_A) S^-_B (\tau_B) + {\rm h.c.}] \right>$, where $\mathcal{T}$ is the time-ordering operator.  These extra operators $ S^+_A(\tau_A), S^-_B (\tau_B)$ are referred to as worm operators.  A worm algorithm is an algorithm in which valid configurations for the partition function are strongly modified by the following procedure (respecting detailed balance for every update): First, a worm pair is inserted in an existing configuration; second, (one of) these worms (is) are shuffled around in imaginary time and over the entire lattice; and third, the worm pairs are removed  when they are again close to each other. This results in a nearly totally decorrelated new configuration. The algorithm is known to be particularly successful for models with $U(1)$ or $SU(2)$ symmetry breaking with short-ranged interactions.
The implementation here is based on Ref.~\cite{Sadoune2022}, with two small changes: (i) Our potential energy terms, $H_{\rm Z}$ are purely local, resulting in a very fast evaluation of them; (ii) the long-range nature of the spin-exchange terms means that updates which insert or remove such a spin-exchange term must be modified compared to the nearest-neighbor case. We found it convenient to compute the sum of the dipolar potential over the lattice, $\mathcal{N} = \sum_{( {i}, {j} ) } \frac{1}{r_{ij}^3}$, and consider the normalized dipolar potential a probability distribution from which to select the site $j$ given $i$. This leads to a very fast algorithm, in which large system sizes $L \sim 100-200$ can straightforwardly be simulated. The presence of a dipolar interaction term  $S^z_{i}S^z_{j}$ would however lead to a serious reduction in accessible system sizes and also cause a substantial slowdown because any evaluation of the potential energy term involves a sum over the entire lattice.   \\

The QMC calculations use periodic boundary conditions unless we study the experimental setup. Lattice distances in this case are always understood as follows: In order to compute the distance between two sites at $\mathbf{r}_i$ and $\mathbf{r}_j$, one computes $(r_{j,x} - r_{i,x})$ and if this difference is larger than $L/2$ or smaller than $-L/2$, then one shifts by $L$ in order to bring the separation back into the interval $]-L/2, L/2]$. The same procedure is applied along the $y$-axis. \\

Because the dipolar potential decays slowly on a two-dimensional lattice, finite size effects are expected to be important, especially at low temperatures. It is often argued in the literature~\cite{Luijten2002, Fukui2009} to replace the dipolar potential by an Ewald-like resummed potential, in which one sums over periodic images. Specifically,
\begin{equation}
V^{\rm resum}({ \mathbf r} = (x,y) ) = \sum_{n,m} V(x + n L, y + m L).
\end{equation}
In order to compare correlation functions with experiment, we always worked with the bare potential. For the determination of critical properties we found it also convenient  to work with the bare potential as we found the interpretation of the correlation functions more straightforward. However, for properties at low temperature where the system is far more sensitive to the tail of the potential, the resummed potential provides a substantial advantage in reducing the finite size effects. Therefore,  in the entropy calculation (Fig.~4 of the main text) we resorted to the resummed potential.

\subsection{Low temperature analysis}

\begin{figure}
    \includegraphics[width=\columnwidth]{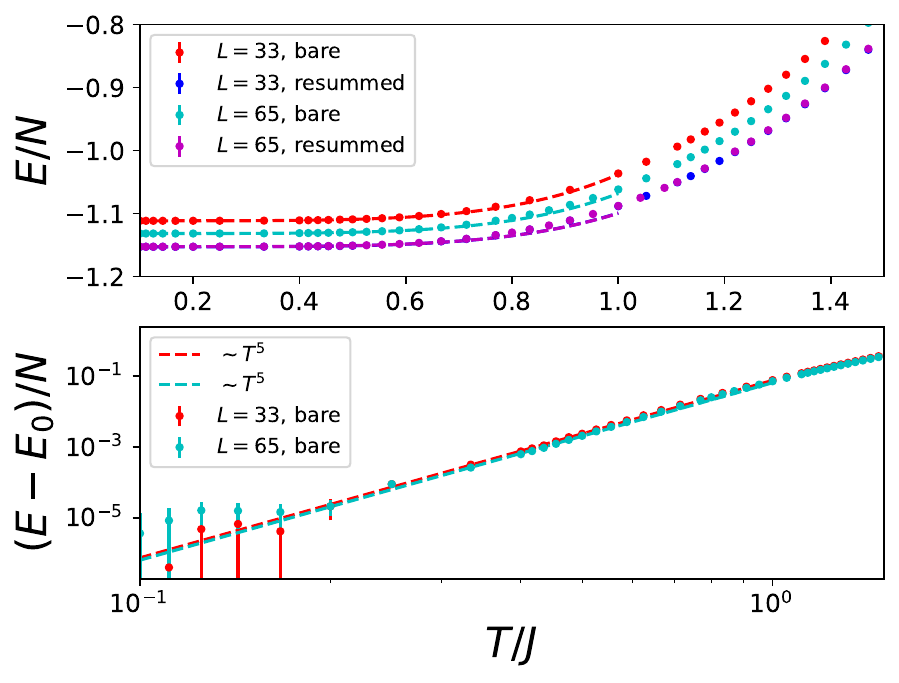}
    \caption{Upper: Energy per site as a function of temperature $T/ J$ for a system with periodic boundary conditions, showing strong finite-size effects. The energy is fitted against the predictions of Eq.~\eqref{eq:E_lowT} (dashed lines). Lower: At low temperatures, the energy difference with the ground state energy is given by a form predicted by linear spin wave theory (dashed line, $\sim T^5$). It remains valid up to $T/J \approx 0.8$, although the fits work best up to $T/J \approx 0.4$.
    }
    \label{fig:QMC_energy_lowT}
\end{figure}

At low temperature, the FM system shows true long-range order with in-plane magnetic XY order. The lowest excitations are spin waves with dispersion $\epsilon = c \sqrt{k} $ for a sufficiently large system \cite{Peter2012}. On dimensional grounds, this leads to the following form for the energy per site
\begin{equation}
\frac{E(T \to 0)}{N} = \frac{E_0}{N} + \gamma T^5 + \ldots.
\label{eq:E_lowT}
\end{equation}
This is confirmed from QMC in the top panel of  Fig.~\ref{fig:QMC_energy_lowT}, where we also observe strong finite size effects on $E_0$. For comparison, note that the energy for the $L=65$ system with the resummed potential is only $3\%$ lower than the energy of a product state with all spins oriented in the $x-$direction. The $T^5$ power-law of the spin waves is seen on the log-log plot of Fig.~\ref{fig:QMC_energy_lowT} (bottom) for systems with linear system size $L=33$ and $L=65$ with periodic boundary conditions, after shifting the energy by the fitted ground state energy. Our fitted values of $\gamma$ ($\gamma = 0.047$ for $L=65$, resummed potential, and $\gamma = 0.0594$ for $L=65$, bare potential) are close but deviate slightly from the prediction of spin wave theory ($\gamma = 0.03935$). This is because of the remaining finite size effects, uncertainties about the fitting interval, and a strong covariance between $\gamma$ and the unknown ground state energy. With a free $\gamma$, the $T^5$ behavior remains valid up to $T/J \approx 0.8$, although the fits are best stabilized up to $T/J =0.4$. The entropy (per particle) of the spin waves is given by $S(T) = 5/4 \gamma T^4$, and remains hence very low, even at $T/J \approx 1$, as could have been expected: the dipolar interactions modify the dispersion of the spin waves (compared to short-range models) in such a way that the energy of the ground state wins over the proliferation (entropy) of the spin waves. Note that uncertainties in $\gamma$ do not strongly affect the entropy at higher temperatures, given our resolution of several percent on the entropy, and our cutoff of the low-energy theory at $T/J = 0.5$.

\subsection{Critical behavior}

\begin{figure}
    \includegraphics[width=\columnwidth]{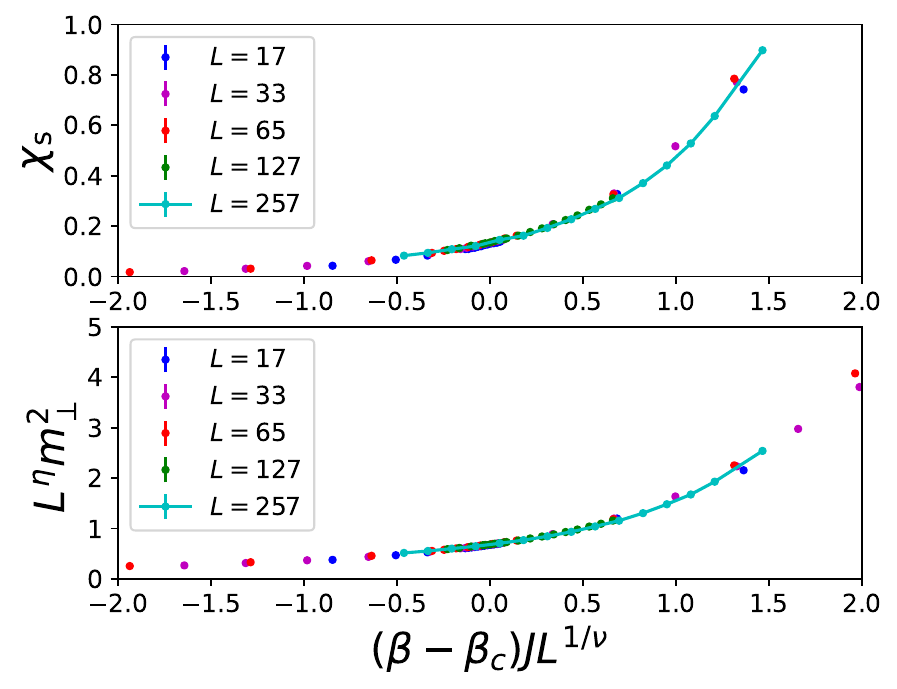}
    \caption{Lower: Data for the rescaled in-plane magnetization squared can be collapsed onto a single curve in the critical region for a system with $\delta=0$ in the vicinity of the critical temperature. Critical exponents are $\nu = 1$ and $\eta=1$, and the applied inverse critical temperature is $\beta_c = 0.5198$. Upper: Collapse of the spin stiffness for the same system.
    }
     \label{fig:QMC_collapse_finiteT}
\end{figure}

\begin{figure}
    \includegraphics[width=\columnwidth]{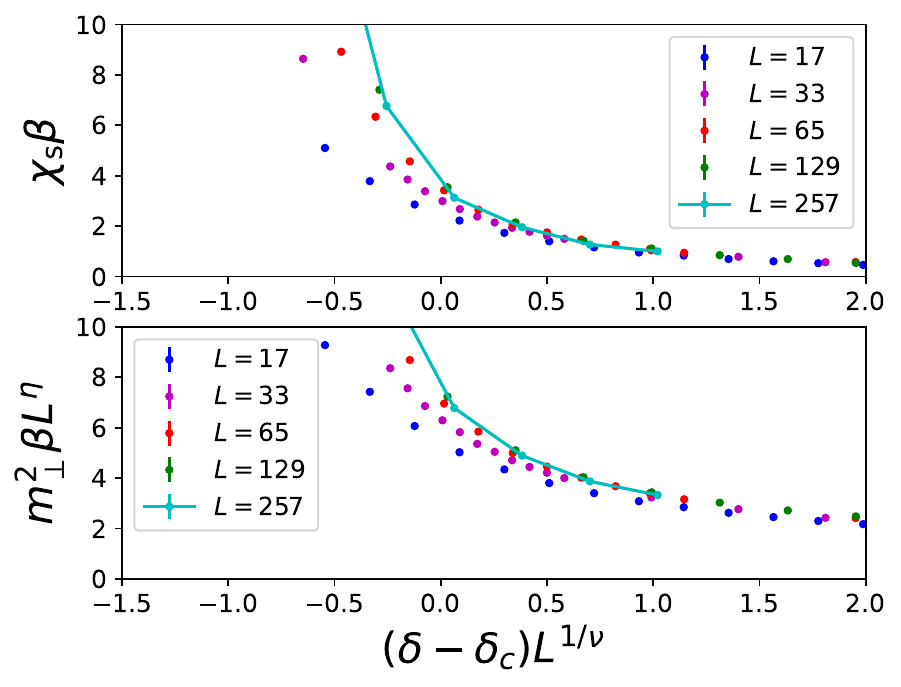}
    \caption{Lower: Data for the rescaled in-plane magnetization squared can be collapsed onto a single curve in the critical region in the vicinity of the quantum critical point $\delta_c = 4.03$. Critical exponents are $z=1/2$, $\nu = 1$ and $\eta=1$. Upper: Collapse of the spin stiffness for the same system. 
    }
     \label{fig:QMC_collapse_zeroT}
\end{figure}

We provide additional plots in order to discuss the critical behavior of the second order phase transitions. In Fig.~\ref{fig:QMC_collapse_finiteT} it is shown that with the right critical exponents $\eta=1$ and $\nu=1$, the curves of the (scaled) in-plane magnetization squared $m_{\perp}^2$  and the spin stiffness $\chi_{\rm s}$ can be collapsed onto a single curve as a function of a (scaled) inverse temperature in the vicinity of the critical temperature. For the curves corresponding to the smallest system sizes $L < 65$, small deviations are visible due to finite size effects. 
The same behavior is seen for the quantum phase transition where we scaled the inverse temperature $\beta$ as $\beta = L^z = L^{1/2}$, see Fig.~\ref{fig:QMC_collapse_zeroT}. The finite size effects are more pronounced than for the thermal transition with $\delta=0$; a reasonable collapse is only found for $L \ge 97$.

\subsection{Correlation functions}

\begin{figure}
    \centering
    \includegraphics[width=\columnwidth]{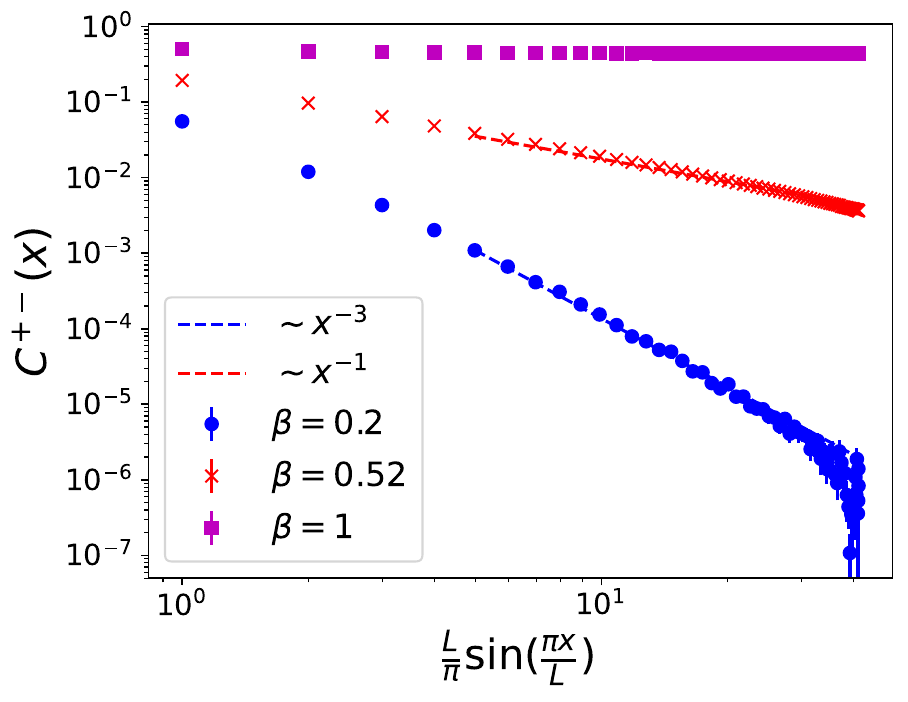}
    \caption{The spin exchange correlation function decays asymptotically with a power-law  with power 3 in the paramagnetic phase, power 1 at the critical temperature, and approaches a constant in the ordered phase. The usage of the cord function on the $x-$axis is convenient to remove the leading order effects of the periodic boundary conditions. Simulations were performed for $\delta = 0$ and $L=129$ with the inverse temperature indicated in the figure. The results are plotted along the $x-$axis only.
    }
     \label{fig:QMC_powerlaw}
\end{figure}

\begin{figure}
    \centering
    \includegraphics[width=\columnwidth]{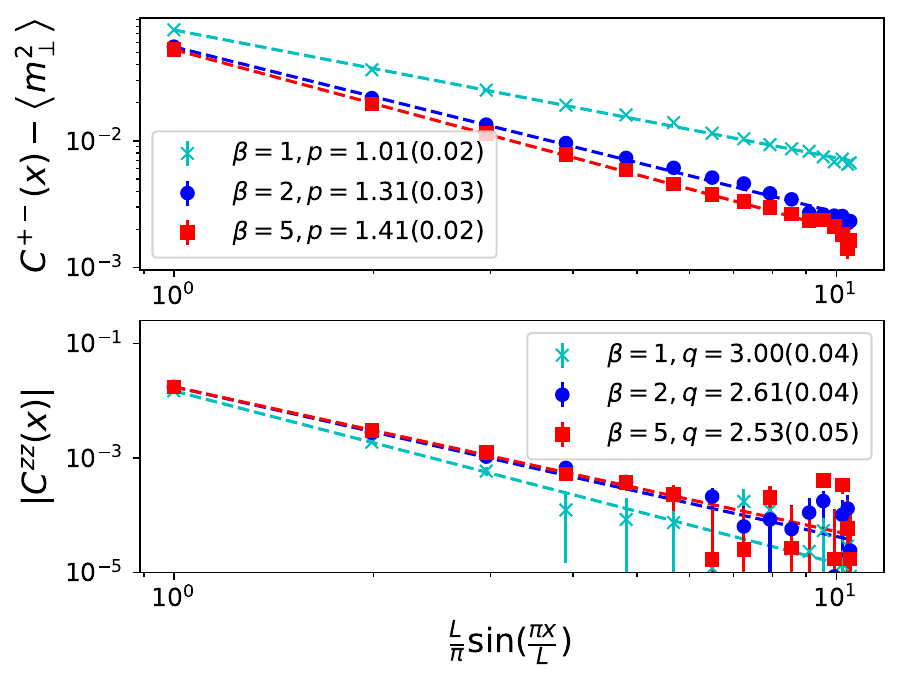}
    \caption{Extracting the decay of the $C^{+-}(x) - \left< m_{\perp}^2 \right> \sim x^{-p}$ and $\vert C^{zz}(x) \vert \sim x^{-q}$ correlation functions in the ordered phase on approach to the ground state for a system of linear system size $L=33$, with periodic boundary conditions and making use of the bare potential. 
    }
     \label{fig:QMC_powerlaw_gs}
\end{figure}

Ref.~\cite{Peter2012} makes a prediction for the decay of the spin-exchange correlation function $C^{+-}(x)$ in the paramagnetic, critical, and ferromagnetic phases. In the paramagnetic phase, the decay is a power-law with exponent 3 (the same as the dipolar term in the Hamiltonian). At the critical temperature, the decay is asymptotically $1/x$ (as expected from the fact that we established $\eta = 1$ already). In the ordered phase, the spin-exchange correlation function approaches a constant. We verified this in Fig.~\ref{fig:QMC_powerlaw}. 
In the ordered phase, the constant is approached as a power-law, $C^{+-}(x) - m_{\perp}^2 \sim x^{-p}$, with $p = 1.0$ at $0 < T < T_c$ and $p=1.5$ at $T=0$, according to Ref.~\cite{Peter2012}. This is reflected in our data shown in the upper panel of Fig.~\ref{fig:QMC_powerlaw_gs}. We are restricted to a linear system size $L=33$ for this calculation, but it turns out that this suffices to extract $p$.
The case $\beta=1$ is outside the regime where linear spin wave theory is valid and we clearly extract the power $p=1.01(2)$, agreeing with Ref.~\cite{Peter2012} within error bars. For the case $\beta = 5$ we are close to the ground state, and this is reflected in a power, $p=1.41(2)$, a value which is almost $1.5$ and certainly different from $p=1$. Our data for $\beta=10$ (not shown) are very close to the data for $\beta=5$. Considerably lower error bars would be needed before we can make use of the data for $\beta=10$, but we expect a value for $p$ that converges towards $1.5$ with increasing $\beta$. For the case $\beta = 2$, the extracted power-law  $(p=1.31)$ falls in between the finite temperature and ground state predictions. This reflects that on a finite system convergence towards the ground state is faster for small distances than for large distances.

 Spin correlations along the $z$-direction are measured by the correlation function $C^{zz}(x) = \left< S^z(x)S^z(0) \right>$. The magnitude of the values of this correlation function, is very small, given the absence of such interactions in the Hamiltonian. The prediction of Ref.~\cite{Peter2012} is $\vert C^{zz}(x) \vert \sim x^{-q}$, with $q = 3$ for $0 < T < T_c$ and $q = 5/2$ at $T=0$. This is also confirmed in the lower panel of Fig.~\ref{fig:QMC_powerlaw_gs}, where $\beta = 1$ lies again clearly in the $0 < T < T_c$ regime, and $\beta = 5$ close to the ground state. 
At higher temperature, the signal of $C^{zz}(x)$ is too small compared to the statistical noise, so the asymptotic decay of $C^{zz}(x)$ at $T_c$ and in the paramagnetic phase cannot be tested, although the correctness of Ref.~\cite{Peter2012} is beyond doubt.

\section{Pseudo-Majorana functional renormalization group (pm-fRG)}
\label{app:pm-fRG}

\subsection{Pseudo-Majorana representation and vertex functions}
The pm-fRG developed in Refs.~\cite{niggemannFrustratedQuantum2021,niggemannQuantitativeFunctional2022} builds on the faithful representation of spin-1/2 operators in terms of three Majorana fermion operators, $S_j^x=-i\eta_j^y\eta_j^z$, $S_j^y=-i\eta_j^z\eta_j^x$ and $S_j^z=-i\eta_j^x\eta_j^y$ \cite{tsvelikNewFermionic1992}. The Majorana operators fulfill $\{ \eta_{i}^{\alpha},\eta_{j}^{\beta}\} =\delta_{\alpha\beta}\delta_{ij}$, $(\eta_{i}^{\alpha})^{2}=1/2$ and $\eta_{i}^{\alpha}=(\eta_{i}^{\alpha})^\dagger$ where $\alpha,\beta\in{x,y,z}$. Spin Hamiltonians  with bilinear spin-interactions (like $H_\mathrm{XY}$) are thus mapped to purely interacting Majorana systems which are then treated by the functional renormalization group \cite{metznerFunctionalRenormalization2012}. The pm-fRG is similar to the previously established $T=0$ pseudo-fermion fRG \cite{reutherJ1J2Frustrated2010} which, however, builds on a complex-fermionic spin representation. However, the latter introduces non-physical states in the Hilbert space causing a number of problems, see Ref.~\cite{schneiderTamingPseudofermion2022a}. \\

The central building block of the pm-fRG are 2- and 4-point Majorana imaginary time-ordered correlation functions. Their Fourier transforms can be expressed via the Grassmann functional field integral formalism, e.g.~
\begin{equation}
    \bigl\langle\zeta_{i_{1}}^{\alpha_{1}}(\omega_{1})\zeta_{i_{2}}^{\alpha_{2}}(\omega_{2})\bigr\rangle=T\delta_{\omega_{1},-\omega_{2}}\delta_{i_{1},i_{2}}\delta_{\alpha_{1},\alpha_{2}}G_{i_{1}\alpha_{1}}(\omega_{1}).
    \label{eq:pm-corr}
\end{equation}
Here, $\zeta_i^\alpha$ is the (anti-commuting) Grassmann field corresponding to the Majorana operator $\eta_i^\alpha$ and $\omega_{1,2}$ are fermionic Matsubara frequencies. The Kronecker-deltas on the right hand side follow from imaginary-time translation symmetry, the local $\mathbb{Z}_2$ gauge symmetry of the pseudo-Majorana representation ($\eta_i^\alpha\rightarrow-\eta_i^\alpha\;\forall\,\alpha$), and time-reversal symmetry, respectively. While Eq.~\eqref{eq:pm-corr} defines the (full) Majorana propagator $G_{i_{1}\alpha_{1}}(\omega_{1})$, the one-line irreducible self-energy $\Sigma_{j,\alpha}(\omega)=i\gamma_{j,\alpha}(\omega)$ and the one-line irreducible vertex $V_{i\alpha_{1},i\alpha_{2},j\alpha_{3},j\alpha_{4}}(\omega_{1},\omega_{2},\omega_{3},\omega_{4})\equiv V_{ij;\alpha_{1}\alpha_{2};\alpha_{3}\alpha_{4}}\left(s,t,u\right)$ are defined via the tree expansion \cite{niggemannFrustratedQuantum2021,niggemannQuantitativeFunctional2022}. For the vertex, it is convenient to switch to combined bosonic Matsubara frequencies $s=\omega_1+\omega_2$, $t=\omega_1+\omega_3$ and $u=\omega_{1}+\omega_{4}$.

\subsection{pm-fRG flow equations}
To implement the renormalization group idea in an exact way \cite{kopietzIntroductionFunctional2010}, the pm-fRG introduces a multiplicative Matsubara frequency cutoff dependent on a scale $\Lambda$ in the bare propagator, $G^{(0)}_{i,\alpha}(\omega)=1/(i\omega) \rightarrow G^{(0),\Lambda}_{i,\alpha}(\omega) = \vartheta_\Lambda(|\omega|)/(i\omega)$. We take $\vartheta_\Lambda(|\omega|)=\omega^2/(\omega^2+\Lambda^2)$, but we checked that the results do not depend significantly on this particular choice. When $\Lambda\rightarrow\infty$, the bare propagator vanishes $G^{(0),\Lambda}\rightarrow 0$. Then the vertices are trivial and frequency independent with $V_{ij;xz;xz}=V_{ij;yz;yz}\rightarrow -J_{ij}$ as the only non-zero contribution. 
\\

We use lattice symmetries (translation and point-group) and $\mathrm{U(}1)$ spin rotation symmetry to reduce the number of independent components of the vertex. However, the list of independent vertices ($V_{ij;xx;xx}$, $V_{ij;zz;zz}$, $V_{ij;xx;yy}$, $V_{ij;xx;zz}$ $V_{ij;xy;xy}$ and $V_{ij;xz;xz}^\Lambda$) is still significantly larger than in the case with full Heisenberg $\mathrm{SO}(3)$ spin rotation symmetry considered so far within pm-fRG~\cite{niggemannFrustratedQuantum2021,niggemannQuantitativeFunctional2022,Niggemann2022}.
\\

The pm-fRG proceeds in lowering the cutoff $\Lambda$ to zero where the original theory is recovered. One-loop flow equations describe the $\Lambda$ dependence of the self-energy $\gamma_{i,\alpha}^\Lambda$ and vertex functions $V^\Lambda$. To write them in compact form, we define $G_{i\alpha}^{\Lambda}(\omega)=-ig_{i\alpha}^{\Lambda}(\omega)$ and parameterize
$g_{i\alpha}^{\Lambda}(\omega)=[\omega/\vartheta_{\Lambda}(\omega)+\gamma_{i\alpha}^{\Lambda}(\omega)]^{-1}$. In addition, the single-scale propagator \cite{kopietzIntroductionFunctional2010} is given by 
\begin{equation}
    \dot{g}_{i\alpha}^{\Lambda}=\left[g_{i\alpha}^{\Lambda}\left(\Omega\right)\right]^{2}[\frac{\Omega\vartheta_{\Lambda}^{\prime}\left(\Omega\right)}{\vartheta_{\Lambda}^{2}\left(\Omega\right)}-\partial_{\Lambda}\gamma_{i\alpha}^{\Lambda}\left(\omega\right)].
    \label{eq:gd_Kat}
\end{equation}
The last term in Eq.~\eqref{eq:gd_Kat} takes into account certain contributions generated by the flowing 6-point vertex (Katanin truncation), but the omission of the remaining parts of the $n-$point vertex at level $n=6,8,10,...$ constitutes the main approximation involved in the practical application of pm-fRG. This approximation is rigorously justified in the perturbative regime $T/|J|\gg 1$ \cite{schneiderTamingPseudofermion2022a}, but for small $T$ care is needed to gauge the validity of the results, see below.
\\

\begin{widetext}
The flow equation for the free energy per spin $f$ \cite{kopietzIntroductionFunctional2010} involves only 2-point quantities ($g_{i\alpha}, \gamma_{i\alpha}$),
\begin{equation}
    \partial_{\Lambda}f^\Lambda=-\frac{T}{2}\sum_{\Omega}\vartheta_{\Lambda}^{\prime}\left(\Omega\right)\vartheta_{\Lambda}^{-1}\left(\Omega\right)\left[2g_{ix}^{\Lambda}\left(\Omega\right)\gamma_{ix}^{\Lambda}\left(\Omega\right)+g_{iz}^{\Lambda}\left(\Omega\right)\gamma_{iz}^{\Lambda}\left(\Omega\right)\right],
    \label{eq:Df}
\end{equation}
and the initial condition is $f^{\Lambda\rightarrow \infty}=-T\,\mathrm{log}(2)$ \cite{niggemannFrustratedQuantum2021}. The flow equations for the self-energies with trivial initial conditions are
\begin{eqnarray}
\partial_{\Lambda}\gamma_{ix}^{\Lambda}(\omega) \!& = & \!-\frac{T}{2}\sum_{\Omega}\sum_{j}\dot{g}_{jx}^{\Lambda}\left(\Omega\right)\left[V_{ji;xx,xx}^{\Lambda}\left(0,\Omega-\omega,\Omega+\omega\right)+V_{ji;xx,yy}^{\Lambda}\left(0,\Omega-\omega,\Omega+\omega\right)\right] \\
 &  & +\dot{g}_{jz}^{\Lambda}\left(\Omega\right)V_{ji;zz,xx}^{\Lambda}\left(0,\Omega-\omega,\Omega+\omega\right)\nonumber,\\
\partial_{\Lambda}\gamma_{iz}^{\Lambda}(\omega) \!& = & \!-\frac{T}{2}\sum_{\Omega}\sum_{j}2\dot{g}_{jx}^{\Lambda}\left(\Omega\right)V_{ji;xx,zz}^{\Lambda}\left(0,\Omega-\omega,\Omega+\omega\right)+\dot{g}_{jz}^{\Lambda}\left(\Omega\right)V_{ji;zz,zz}^{\Lambda}\left(0,\Omega-\omega,\Omega+\omega\right).
\end{eqnarray}
The flow of the 4-point vertices is 
\begin{equation}
    \partial_{\Lambda} V_{ij;\alpha_{1}\alpha_{2};\alpha_{3}\alpha_{4}}^{\Lambda}\left(s,t,u\right)=X_{ij;\alpha_{1}\alpha_{2};\alpha_{3}\alpha_{4}}^{\Lambda}\left(s,t,u\right)-\tilde{X}_{ij;\alpha_{1}\alpha_{3};\alpha_{2}\alpha_{4}}^{\Lambda}\left(t,s,u\right)+\tilde{X}_{ij;\alpha_{1}\alpha_{4};\alpha_{2}\alpha_{3}}^{\Lambda}\left(u,s,t\right)
\end{equation}
where we abbreviated the first contribution involving a lattice sum over sites $k$,
\begin{eqnarray}
X_{ij;\alpha_{1}\alpha_{2};\alpha_{3}\alpha_{4}}^{\Lambda}\left(s,t,u\right) & = & -T\sum_{\Omega}\sum_{\beta_{1,2}}\sum_{k}\dot{g}_{k,\beta_{1}}^{\Lambda}(\Omega)g_{k,\beta_{2}}^{\Lambda}(\Omega+s)\\
 &  & \times V_{ik;\alpha_{1}\alpha_{2};\beta_{2}\beta_{1}}^{\Lambda}\left(s,-\Omega-\omega_{2},\omega_{1}+\Omega\right)V_{kj;\beta_{1}\beta_{2};\alpha_{3}\alpha_{4}}^{\Lambda}\left(s,-\Omega+\omega_{3},-\Omega+\omega_{4}\right).\nonumber
\end{eqnarray}
The remaining terms are given by $\tilde{X}_{ii;\alpha_{1}\alpha_{2};\alpha_{3}\alpha_{4}}^{\Lambda}\left(s,t,u\right)=X_{ii;\alpha_{1}\alpha_{2};\alpha_{3}\alpha_{4}}^{\Lambda}\left(s,t,u\right)$ for the on-site ($ii$) "bond". For non-trivial bonds, $i\neq j$, there is no internal lattice sum in $\tilde{X}$:
\begin{eqnarray}
\tilde{X}_{ij;\alpha_{1}\alpha_{2};\alpha_{3}\alpha_{4}}^{\Lambda}\left(s,t,u\right) & = & T\sum_{\Omega}\sum_{\beta_{1,2}}\left[\dot{g}_{i,\beta_{1}}^{\Lambda}(\Omega)g_{j,\beta_{2}}^{\Lambda}(\Omega+s)+g_{i,\beta_{1}}^{\Lambda}(\Omega)\dot{g}_{j,\beta_{2}}^{\Lambda}(\Omega+s)\right]\\
 &  & \times V_{ij;\alpha_{1}\beta_{1};\alpha_{2}\beta_{2}}^{\Lambda}\left(\omega_{1}+\Omega,s,-\Omega-\omega_{2}\right)V_{ij;\beta_{1}\alpha_{3};\beta_{2}\alpha_{4}}^{\Lambda}\left(-\Omega+\omega_{3},s,-\Omega+\omega_{4}\right).\nonumber
\end{eqnarray}
The $\beta_{1,2}$-sums run over the Majorana flavors $x,y,z$ and the required combinations of $\alpha_1\alpha_2;\alpha_3\alpha_4$ identified above are yet to be inserted. The numerical complexity is reduced by a number of symmetries related to the anti-commutation relation of Grassmann fields \cite{niggemannQuantitativeFunctional2022}. 
\\

Finally, at the end of the flow $\Lambda\rightarrow0$, physical observables like spin susceptibilities $\chi_{ij}^{\alpha\alpha}(i\nu)=\left\langle S_{i}^{\alpha}(i\nu)S_{j}^{\alpha}\right\rangle$ can be computed, e.g.~
\begin{eqnarray}
    \chi_{ij}^{xx}(i\nu)&=&\delta_{ij}T\sum_{\omega}g_{ix}\left(\omega\right)g_{iz}(\omega-\nu)\label{eq:chi_xx}\\
    & &+T^{2}\sum_{\omega,\omega^{\prime}}g_{ix}\left(\omega\right)g_{iz}\left(\omega-\nu\right)g_{jx}\left(\omega^{\prime}+\nu\right)g_{jz}\left(\omega^{\prime}\right)V_{ij;yz;yz}\left(\nu,\omega-\omega^{\prime}-\nu,\omega+\omega^{\prime}\right)\nonumber.
\end{eqnarray}
Equal-time correlators can then be obtained from $\chi_{ij}^{xx}(i\nu)$ via a sum over $\nu$,
\begin{equation}
    C_{ij}^{xx}=\left\langle S_{i}^{x}S_{j}^{x}\right\rangle =\frac{\delta_{ij}}{4}+T^{3}\sum_{\Omega}\sum_{\omega,\omega^{\prime}}g_{ix}\left(\omega\right)g_{iz}\left(\omega-\Omega\right)g_{jx}\left(\omega^{\prime}+\Omega\right)g_{jz}\left(\omega^{\prime}\right)V_{ij;xz;xz}\left(\Omega,\omega-\omega^{\prime}-\Omega,\omega+\omega^{\prime}\right).\label{eq:Cijxx_pm-fRG}
\end{equation}
For further details on the numerical implementation, we refer to Refs.~\cite{niggemannFrustratedQuantum2021,niggemannQuantitativeFunctional2022}.
\end{widetext}

\subsection{Consistency checks for the AFM case}
As mentioned above, it is important to critically assess the validity of the pm-fRG results in the low-temperature regime. For the FM case, as explained in the main text, we have directly compared the pm-fRG against various results of the error-controlled QMC, see e.g.~$C_{ij}^{xx}$ in Fig.~1. 
For the AFM case, beyond comparison to HTE for large $T$ (Fig.~\ref{fig:QMC_energy_highT}) no independent benchmarks are available and we have to resort to method-intrinsic consistency checks. 
\begin{figure*}
    \centering
    \includegraphics{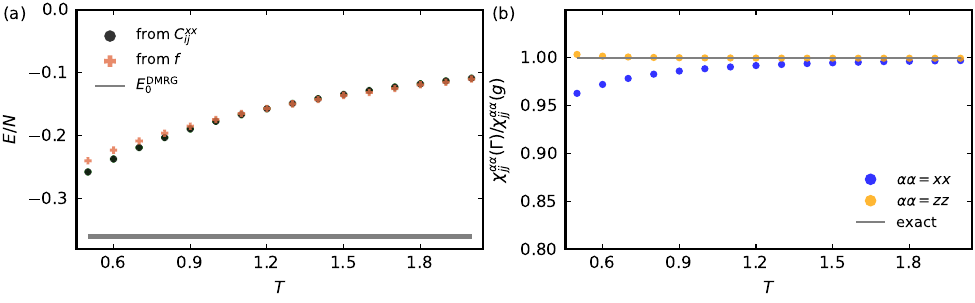}
    \caption{Benchmarking the pm-fRG on the AFM side ($J=1$): (a) Energy per spin over temperature as calculated from pm-fRG, both via $\langle H_\mathrm{XY} \rangle$ from equal time correlators $C_{ij}^{xx}$ (dots) and from $E/N=\partial_\beta(f\beta)$  (crosses) where $f$ is the free energy per spin. The horizontal line denotes the DMRG ground state energy reproduced from Ref.~\cite{chen_continuous_2022}.
    (b) Ratio of local spin susceptibilities (at $\nu=0$) calculated via the 4-point pseudo-Majorana vertex $\Gamma$ and directly from the Majorana propagator $g$. We show the susceptibility-ratio both for in-plane ($x$) and perpendicular ($z$) direction.}
     \label{fig:AFM_u}
\end{figure*}

A first check involves the energy per spin $E/N$. This quantity can be calculated as $\langle H_\mathrm{XY} \rangle/N$ based on the equal time correlators of Eq.~\eqref{eq:Cijxx_pm-fRG} which depend directly on the 4-point vertices $V$, see black dots in Fig.~\ref{fig:AFM_u}(a). The energy can as well be calculated from the derivative of the free energy, $E/N=\partial_\beta(f\beta)$ (brown crosses) where $f$ is obtained from its own flow-equation \eqref{eq:Df} which depends on $V$ only indirectly via its appearance in the self-energy flows. Thus, an agreement between both ways to calculate the energy signals an internal consistency between 2- and 4-point objects. The data shown in Fig.~\ref{fig:AFM_u}(a) reveals good agreement between both approaches. The difference grows as $T$ is lowered, reaching $7\%$ at $T=0.5$. We suggest that this strategy is generally applicable to gauge the validity of fRG simulations beyond the pm-fRG.
\\

A similar but independent check specific to pm-fRG builds on the fact that for any spin-1/2 Hamiltonian written in terms of pseudo-Majoranas, the operator $\Theta_{j}\equiv-2i\eta_{j}^{x}\eta_{j}^{y}\eta_{j}^{z}$ is a constant of motion, i.e.~$\Theta_{j}(\tau)=\Theta_{j}$. Using $\Theta_{j}^2=1/2$, one finds that the {\it local} spin susceptibility can be calculated as
\begin{equation}
   \chi_{jj}^{xx}(i\nu=0)=\sum_{n}\frac{1}{\pi\left(2n+1\right)}g_{jx}\left(\omega_{n}\right), \label{eq:chi_xx_COM}
\end{equation}
where we focused on the static part. Again, this expression does not involve the vertex $V$ directly, in contrast to the general expression \eqref{eq:chi_xx}. The ratio of $\chi_{jj}^{xx}(i\nu=0)$ calculated either by Eq.~\eqref{eq:chi_xx} or Eq.~\eqref{eq:chi_xx_COM} is shown in Fig.~\ref{fig:AFM_u}(b). Its deviation from unity represents a measure for the violation of the constant-of-motion property and signals the degree of internal inconsistency between 2- and 4-particle objects caused by the fRG truncation. Again, the ratio grows with decreasing $T$ and for $T=0.5$ deviates from unity by $4\%$. In other models with available benchmarks like small spin clusters, deviations of this size have been found associated to few-percent errors in common observables like susceptibilities or equal-time correlators. 
\\

Comparing the outcome of these consistency checks to other models where exact benchmark data is available (FM-side, small spin clusters) suggests that the pm-fRG results for the AFM case are to be trusted within a few-percent error margin for the reported temperature range $T\geq 0.5J$. Note however that even a perfect consistency between 2- and 4-point vertices would not be a guarantee for the exactness of a many-body calculation. For example, such a consistency is a general feature of conserving approximations.
\\

Finally, via variation of the maximally allowed spatial range of the 4-point Majorana vertex in pm-fRG, the actual spin-spin correlation length can be estimated. The data in Fig.~1 of the main text are obtained by cutting off the Majorana vertex function $V_{ij}$ for $|\mathbf{r}_i - \mathbf{r}_j|>16$, but the results for susceptibilities do not change when this range is lowered to $10$. This is in contrast to the finite-size scaling behavior that is found for the FM side in the vicinity of $T_c$ (data not shown). We thus conclude that for the AFM case, (i) the pm-fRG data presented are essentially for an infinite system and (ii) the lowest temperature $T=0.5$ achievable in pm-fRG is still above the critical temperature for the anticipated Kosterlitz-Thouless transition which would be signaled by a diverging correlation length \cite{dingPhaseTransition1992}.

\bibliography{dipolarXY}

\end{document}